# Confidence intervals with a priori parameter bounds

A.V. Lokhov*, F.V. Tkachov

Institute for nuclear research RAS, Moscow, 117312

*e-mail: lokhov.alex@gmail.com

*Abstract.* We review the methods of constructing confidence intervals that account for *a priori* information about one-sided constraints on the parameter being estimated. We show that the so-called method of sensitivity limit yields a correct solution of the problem. Derived are the solutions for the cases of a continuous distribution with non-negative estimated parameter and a discrete distribution, specifically a Poisson process with background. For both cases, the best upper limit is constructed that accounts for the a priori information. A table is provided with the confidence intervals for the parameter of Poisson distribution that correctly accounts for the information on the known value of the background along with the software for calculating the confidence intervals for any confidence levels and magnitudes of the background (the software is freely available for download via Internet).

## 1. Introduction

The Neyman construction [1] of confidence intervals for estimated parameters is a basic element of experimental data processing. Often one also possesses a priori information about the estimated parameters, and it is important to include that information into the confidence intervals in a consistent way.

A limited domain of the parameters is an example of such a priori information. The problem with the conventional confidence intervals is seen if the experimental estimate of the parameter falls out of the domain. For instance, in the Troitsk-nu-mass experiment on the direct measurement of the mass of neutrino in tritium beta-decay [2] the parameter $m_\nu^2$ is non-negative while the formal fit yields a negative value of $m_\nu^2$.

The construction of confidence intervals for Poisson distribution with Poisson-distributed background is another situation where one should take into account the a priori information about the background. The situation is usual for studying rare events (in experiments on neutrinoless double beta-decay [3], and neutrino oscillations, for instance T2K, MINOS [4], etc.)

Several candidate solutions were proposed. These can be divided into two groups according to how the freedom inherent in the Neyman construction of confidence intervals is used.

The first group of candidate solutions incorporates the a priori information at the stage of constructing the acceptance regions. This group includes the Feldman-Cousins construction [6] and the power constrained limits advocated by Cowan et al. (the CCGV method [7]). However, the intervals constructed e.g. via the Feldman-Cousins recipe do not allow one to meaningfully compare the results of different experiments (because an experiment with worse sensitivity could yield a smaller interval), thus failing to achieve the very goal of data processing: to produce numbers that directly express the essential information; numbers that speak for themselves.

The second group of candidate solutions incorporates the a priori information into the estimator, and only after that proceeds to constructing confidence intervals in a regular fashion. The first such recipe was given in [8] for a special case of the maximal likelihood estimator. The somewhat artificial arguments of [8], however, are not quite transparent and no explicit form for the estimator is provided, so that it is not clear whether the recipe could be applied to other methods of estimation (e.g. the least squares or the newer and advantageous method of quasi-optimal weights [17] which was used in obtaining the recent neutrino mass bound [2]). This limitation of the (correct) construction of confidence intervals [8] may explain why it remained unnoticed by the data processing community.

A comprehensive solution that is independent of the estimation method and provides an explicit formula for the estimator, was found in [9] where a transparent graphical and analytical interpretation of the construction was given. For the unphysical values of the estimator, the resulting construction resembles the empirical recipe of the so-called sensitivity limit (for instance, the sensitivity limit was used among other ways to present the results of the Mainz neutrino mass experiment [10]). For the physical values of the estimator, however, the new construction introduces a narrowing of the confidence belt near the boundary of the physical region. We will call the new construction *the method of sensitivity limit*. To avoid confusion, however, it should be emphasized that the sensitivity limit proper is not a confidence interval but a characteristic of a given experiment and the corresponding uncertainties; it can be calculated before measurements. The method of sensitivity limit, on the other hand, provides a system of confidence intervals (a confidence belt) constructed via the Neyman procedure.

A major advantage of the method of sensitivity limit is that it allows a direct comparison of different experiments without recalculations or re-processing of data. For instance, the new results of the Troitsk-nu-mass experiment [2] are compared with the above mentioned Mainz measurement as well as with the results of the first analysis of the Troitsk-nu-mass data [11]. In this regard the method succeeds where the Feldman-Cousins recipe [6] fails: in non-physical range the Feldman and Cousins recipe provides a confidence interval that depends on the experimental value of the estimator; moreover the interval shrinks unnaturally as the estimator values move away from the physical bound into the unphysical region. In fact, the Feldman-Cousins recipe does not provide results that can be compared directly.

Note that within the second approach (incorporation of a priori information into the estimator prior to constructing confidence intervals) one can also construct correct and optimal one-side (upper or lower) limits for the estimated parameters [12].

The case of discrete distributions is another natural extension of the method of [9], [12]; this is the main purpose of the present paper.

First, in Section 2 we recall the Neyman construction and define it in the terms convenient for the further derivation of the method of sensitivity limit. The case of discrete distributions is discussed separately since there one should replace the equalities for the confidence probability of the confidence belt by inequalities due to the discreteness itself. We discuss the construction of symmetric and non-symmetric as well as the construction of Sterne, Crow and Gardner [5].

Sections 3-5 review the constructions of Cowan et al., Feldman and Cousins and Mandelkern and Shultz correspondingly.

In Section 6, following [9], we construct the method of sensitivity limit for the case of continuous distributions, with a priori information about the estimated parameter given by the inequality $\theta \geq 0$. Section 7 considers the discrete case of a Poisson process with unknown $\mu$ but with known Poisson background *b*. Finally, Section 8 provides the best upper limits for continuous and discrete distributions. Section 9 illustrates that one can compare confidence intervals constructed via the method of sensitivity limit in different tritium $\beta$-decay experiments. The conclusions are summarized in Section 10.

This paper considers only the Neyman procedure of the construction of confidence intervals. The intervals provide clear interpretation within the frequentist approach. Although the alternative Bayesian approach can, with some stretching of imagination, be interpreted in terms of statistical ensembles [13], the actual construction of Bayesian intervals by experimenters uses the unknown a priori distribution density function for the parameter. This violates the applicability of the Bayes theorem and makes it difficult to interpret the results. We also do not consider here the so-called *CLs* method [14] of constructing the confidence intervals, since it has no clear interpretation within either frequentist or Bayesian approach even if it is widely used, for instance, in presenting the results of the Higgs boson searches [15].

## 2. Neyman intervals

### a) Continuous distributions

We start from the description of the standard Neyman construction of confidence intervals [1], in which we fix the notation that is used hereinafter.

Let $\hat{\theta}$ be a conventional estimator for the unknown parameter $\theta$, i.e. an estimator constructed without regard for the a priori bound (e.g. obtained via the paradigmatic method of moments [16], [17]).

The random variable $\hat{\theta}$ is a function of a set of experimental data *X*: $\hat{\theta} = \hat{\theta}(X)$. Its probability density $d_\theta(\hat{\theta})$ is parameterized by $\theta$ and is assumed to be known and non-singular as required in the standard Neyman construction of confidence intervals [1]. The density $d_\theta(\hat{\theta})$ incorporates all the information about the experiment (including the estimation method) in regard of the measurement of $\theta$.

Let $\alpha$, $\alpha'$ be small and non-negative. Define $L_\alpha(\theta)$ and $U_{\alpha'}(\theta)$ according to

$$\mathsf{P}\big(-\infty < \hat{\theta} < L_\alpha(\theta)\big) = \alpha, \quad \mathsf{P}\big(U_{\alpha'}(\theta) < \hat{\theta} < +\infty\big) = \alpha'. \tag{1}$$

The probability for the estimator to fall below $L_\alpha(\theta)$ is $\alpha$, above $U_{\alpha'}(\theta)$, $\alpha'$. The $L_\alpha(\theta)$ thus defined corresponds to the $Z_\alpha$ defined in sec. 9.1.1 of [16].

Assuming $L_\alpha(\theta)$ and $U_{\alpha'}(\theta)$ to be invertible functions of $\theta$, Eq. (1) can be rewritten as follows:

$$\mathsf{P}\big(l_\alpha(\hat{\theta}) < \theta\big) = \alpha, \quad \mathsf{P}\big(\theta < u_{\alpha'}(\hat{\theta})\big) = \alpha', \tag{2}$$

where $u_\alpha = U_\alpha^{-1}$, $l_\alpha = L_\alpha^{-1}$. Eq. (2) says that the probabilities for the random variables $l_\alpha(\hat{\theta}) / u_{\alpha'}(\hat{\theta})$ to fall below/above the unknown true value $\theta$ are $\alpha$ / $\alpha'$.

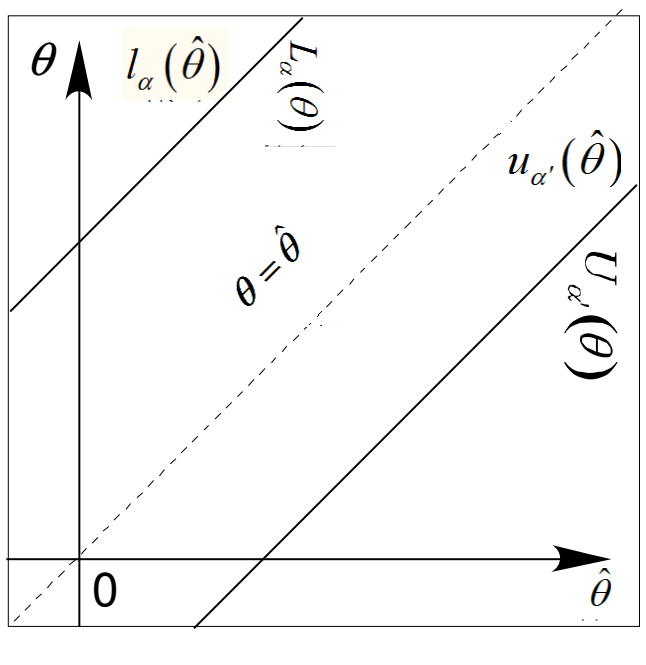

**Fig. 1.** Shown are the functions $\theta = l_\alpha(\hat{\theta})$ and $\theta = u_{\alpha'}(\hat{\theta})$ (or $\hat{\theta} = L_\alpha(\theta)$ and $\hat{\theta} = U_{\alpha'}(\theta)$, depending on the viewpoint). In general, the diagonal $\theta = \hat{\theta}$ need not lie between the two

curves, and it will not be shown in other figures. The two solid curves will be reused in subsequent figures with $\alpha' = \alpha$ , in which case they form the standard symmetric confidence belt for the confidence level $\beta = 1 - 2\alpha$. Smaller $\beta$ means a more narrow belt.

One can rewrite (1) in form of

$$\mathrm{P}\Big(L_\alpha(\theta) < \hat{\theta} < U_{\alpha'}(\theta)\Big) = 1 - \alpha - \alpha' \equiv \beta\,. \tag{3}$$

Then an equivalent expression

$$\mathrm{P}\Big(u_{\alpha'}(\hat{\theta}) < \theta < l_\alpha(\hat{\theta})\Big) = \beta \tag{4}$$

says that the probability for the random interval $[u_{\alpha'}(\hat{\theta}), l_\alpha(\hat{\theta})]$ to cover the unknown $\theta$ is $\beta$ (the confidence level; e.g. $\beta = 90\%$ etc.).

Choosing $\alpha = \alpha' = (1-\beta)/2$ results in a standard symmetric confidence belt.

The construction allows further freedom. Fix the confidence level $\beta$ (e.g. $\beta = 90\%$ ; $\beta$ is assumed to be fixed in what follows). Then choose a pair of functions $L$, $U$ to satisfy

$$\mathrm{P}\Big(L(\theta) < \hat{\theta} < U(\theta)\Big) = \beta\,. \tag{5}$$

If they are also monotonic then there exist inversions $u = U^{-1}, l = L^{-1}$, and an equivalent expression

$$\mathrm{P}\Big(u(\hat{\theta}) < \theta < l(\hat{\theta})\Big) = \beta \tag{6}$$

says that the random interval $\left[u(\hat{\theta}), l(\hat{\theta})\right]$ covers the unknown $\theta$ with probability $\beta$.

Note that the curve $\theta = u(\hat{\theta})$ cannot exceed $\theta = u_{1-\beta}(\hat{\theta})$, i.e. $u(\hat{\theta}) \le u_{1-\beta}(\hat{\theta})$. The curve $\theta = l(\hat{\theta})$ is similarly bounded from below. Any such pair of curves forms what we will call *allowed confidence belt* for the confidence level $\beta$.

Fig. 2 introduces, in addition to the symmetric confidence belt for the confidence level $\beta$, another, narrower symmetric confidence belt for a lower confidence level $\tilde{\beta} = 1 - 2(1-\beta) = 1 - 4\alpha < \beta$ (dashed sloping lines). The various intersection points and horizontal lines are labelled for ease of reference: similarly labelled points in subsequent figures are the same as in this one.

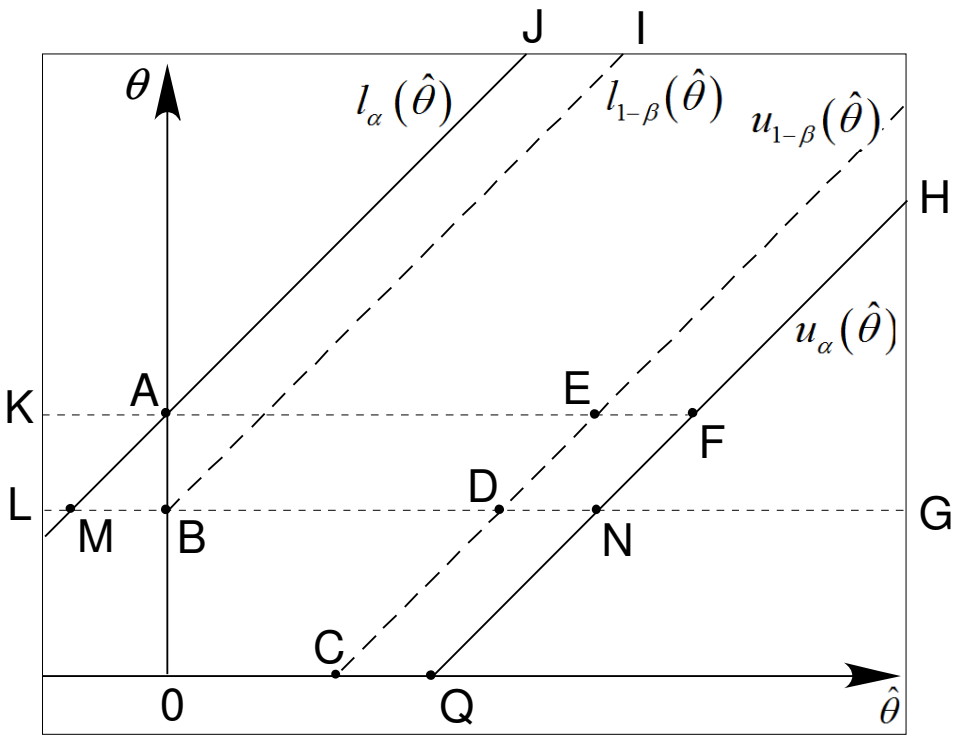

**Fig. 2.** The pairs of solid and dashed sloping lines delimit symmetric confidence belts for the confidence limits $\beta = 1-2\alpha$ and $\tilde\beta = 1-2(1-\beta) = 1-4\alpha$; cf. Fig. 1. The functions that correspond to the lines are shown in the figure. A is the intersection with the vertical axis of the line $\theta = l_\alpha(\hat\theta)$. Point A determines the horizontal line KF along with the further intersection points. C and Q are the intersections of the lines $\theta = u_{1-\beta}(\hat\theta)$ and $\theta = u_\alpha(\hat\theta)$ with the horizontal axis.

The vertical position of the intersection point A (and of the line KF) is denoted as $\theta_{\mathsf{A}}$:

$$\theta_{\mathsf{A}} = l_\alpha(0). \tag{7}$$

The numbers $\theta_{\mathsf{C}} < \theta_{\mathsf{E}} < \theta_{\mathsf{F}}$ are the horizontal positions of the points C, E, F:

$$\theta_{\mathsf{C}} = U_{1-\beta}(0), \quad \theta_{\mathsf{E}} = U_{1-\beta}(\theta_{\mathsf{A}}), \quad \theta_{\mathsf{F}} = U_\alpha(\theta_{\mathsf{A}}). \tag{8}$$

**b) Discrete distributions**

Let us consider an experiment that measures a number of events *n*. Let the number of detected events has, for instance, Poisson distribution:

$$P_\mu(n) = \frac{\mu^n}{n!} e^{-\mu}. \tag{9}$$

Here $\mu$ is the parameter of the Poisson distribution, the mean number of events. (The following reasoning does not depend on the particular type of the distribution.)

The discreteness of the distribution induces some modifications in constructing the confidence intervals as compared with continuous distributions.

As usual, to construct confidence intervals one chooses a confidence level $\alpha$ (95%, for instance). For each value of $\mu$ one can find values n to satisfy the condition:

$$P_\mu\left(n \in \left[n_1(\mu), n_2(\mu)\right]\right) \geq \alpha. \tag{10}$$

The discreteness of the distribution requires weakening the exact equality in **(10)** (as in Eq. **(5)** and **(6)** for continuous distributions) by replacing it with an inequality "$\geq$"; this is the specificity of the discrete distributions already discussed in [6]. Thus, one can find the intervals that will cover the unknown true value $\mu_0$ in fraction not less than $\alpha$ of experiments:

$$P(\mu_0 \in [\mu_1, \mu_2]) \geq \alpha . \quad (11)$$

First, let us consider a one-side interval:

$$P(n \leq n_\alpha(\mu)) \geq \alpha . \quad (12)$$

To derive the confidence interval for the parameter $\mu$ one can perform a transformation of the expression in the brackets:

$$P(\mu \geq \mu^*(n)) \geq \alpha . \quad (13)$$

The inequality (13) implies that in fraction $\geq \alpha$ of experiments (measurements of *n*) one will obtain such values *n* that the unknown true value of the parameter satisfy the condition $\mu \geq \mu^*(n)$.

At first glance it seems that for a discrete distribution of *n* the function $\mu^*(n)$ is defined somehow ambiguously (each value *n* corresponds to a set of values of $\mu^*$ and the values satisfies the condition (13)). As we show below the requirement of conservatism of the intervals [6] (the satisfaction of (13) for each fixed $\mu$) determines the only value of $\mu^*$ for each *n*. The value $\mu^*$ is the (lower) boundary of the confidence interval.

To determine $\mu^*(n)$ we introduce the following notation: let $\mu_n$ be the value of the parameter for which the exact equality $P(n \leq n_\alpha(\mu_n) - 1) = \alpha$ is realized (at that $P(n \leq n_\alpha(\mu_n + \varepsilon) - 1) < \alpha$ and $P(n \leq n_\alpha(\mu_n + \varepsilon)) > \alpha$, where $\varepsilon$ is an arbitrarily small positive quantity).

As we show below if one chooses a value $\mu^*(n) = \mu_n + \varepsilon$ as $\mu^*(n)$ then the conservatism condition is not satisfied.

With such choice of $\mu^*(n)$, if the true value $\mu$ fall into the interval $(\mu_n, \mu_n + \varepsilon)$ the condition (13) will be violated. Indeed in this case $\mu$ will be greater than or equal to a variate $\mu^*(n) = \mu_n + \varepsilon$ only if the measurement yields the values $n = n_\alpha(\mu) - 1$, $n = n_\alpha(\mu) - 2$, etc. The probability to obtain these values is $P(n \leq n_\alpha(\mu > \mu_n + \varepsilon) - 1) < \alpha$ (due to the definition of $\mu_n$ above). Thus, the interval $[\mu_n + \varepsilon, +\infty)$ by definition is not the confidence one.

If one chooses $\mu_n$ as the function $\mu^*(n)$ then the condition $\mu \geq \mu_n$ will be satisfied in a fraction $\geq \alpha$ of measurements and it is in agreement with the definition of the confidence interval for the parameter $\mu$.

Thus, the one-sided confidence interval in the case of a discrete distribution is given by the condition $\mu \geq \mu_n$, where $\mu_n$ is defined by the exact equality $P(n \leq n_\alpha(\mu_n) - 1) = \alpha$.

Similarly, one can consider the one-sided interval defined by the condition $P(n \geq n_\alpha(\mu)) \geq \alpha$. The corresponding upper boundary of the confidence interval for the parameter $\mu$ is defined by the condition $\mu \leq \mu_n'$. Here $\mu_n'$ is given by $P(n \geq n_\alpha(\mu_n') + 1) = \alpha$.

The discreteness of the parameter's distribution introduces additional freedom into the construction of the two-sided confidence belts. For instance, one can specify the two-sided confidence interval as a combination of the upper and the lower boundaries of the one-sided intervals:

$$P(n_1(\mu) \leq n \leq n_2(\mu)) \geq \alpha \quad \Rightarrow P(\mu_1(n) \leq \mu \leq \mu_2(n)) \geq \alpha. \tag{14}$$

The values $\mu_1$ and $\mu_2$ are chosen as $\mu_n$ and $\mu_n'$ correspondingly. The latter values for each $n$ depend on the confidence level $\alpha$.

The choice leads to nearly symmetrical confidence belt. The difference from the case of continuous distributions is that the probabilities contained in the areas $n < n_1(\mu)$ and $n_2(\mu) < n$ can be unequal.

To derive the confidence belt one can also imply some physical reasoning. For instance, if the upper boundary of the interval for the parameter is more important, one can require the condition $P(n < n_1(\mu)) < (1-\alpha)/2$ to be satisfied. Then the area upwards of the upper boundary contains the amount of probability $< (1-\alpha)/2$. The lower boundary of the confidence belt is chosen as the left limit in the condition $P(n_1(\mu) \leq n \leq n_2(\mu)) \geq \alpha$. In this case the condition $P(n > n_2(\mu)) < (1-\alpha)/2$ may not be satisfied and the lower boundary may not coincide with the one derived form (12). At the same time the confidence belt will be less overcovering, its amount of probability can be closer to the desired value $\alpha$ (as it was correctly mentioned in [6] the overcovering is an unwanted quality of the confidence intervals). Similarly one can derive the interval in the case when the lower boundary is more important.

Another way to construct the confidence intervals was suggested by Sterne, Crow and Gardner [5]. Their idea is to construct the acceptance region by adding points into it in the order of ascending probability (as opposed to constructing the symmetrical confidence belt). The method was first applied to the binomial (discrete) distribution, though one can extrapolate it to the case of continuous distributions. The construction generally leads to the asymmetric confidence belt, but the belt has the least possible amount of probability. Crow showed that the method provides the confidence belt with the least possible area.

Hence, for each experimental value of $n$ one can assign a confidence interval for the unknown parameter $\mu$ in the form of

$$\mu \in [\mu_{n\alpha}, \mu'_{n\alpha}]$$

in several different ways. As it is usual in data analysis, it is for an experimenter to decide which method of deriving the confidence belt to choose.

### 3. CCGV construction

Let's turn to the known recipes of taking into account the a priori information while constructing confidence intervals. First we consider the attempts to take the a priori information into account during the construction of the acceptance region (the first group of candidate solutions).

Cowan et al. suggested a method called Power-Constrained limits (PCL) [7]. The idea of the method is as follows. First, one chooses a statistical criteria with statistics $q_\mu = q_\mu(x)$ (it is convenient to use statistics that increases with the growth of the discrepancy between the data $x$ and the value of the parameter $\mu$). Than one derives the power function of the criteria, which is the essence of the whole construction:

$$M_{\mu'}(\mu) = P\left(q_\mu(x) > q_{\mu,crit} \middle| \mu'\right)$$

Than the two hypotheses ($\mu' = 0$ (no signal) and $\mu > 0$ (non-zero signal)) are considered together with the corresponding power function $M_0(\mu)$. One chooses a threshold value of the power function $M_{\min}$ and the range of values of $\mu$ is divided into two. If $M_0(\mu)$ for a value of $\mu$ lies below the threshold than the sensitivity to the parameter is considered to be too low and the values of $\mu$ can not be tested. Thus $\mu$ is not included into the confidence belt for the set of data if 1) the value of $\mu$ is rejected by the criteria $q_\mu$ for the given confidence level $\alpha$, 2) if the sensitivity to the value of $\mu$ is sufficient, i.e. $M_0(\mu) \geq M_{\min}$.

All the values $\mu$ that do not satisfy either condition 1) or 2) form the sought for confidence interval.

The probability of the confidence interval to cover the given $\mu$ equals 100% for the values of $\mu$ for which the power function is below the threshold, and the probability equals $\alpha$ for the values of $\mu$ for which the power function is greater then or equal to the threshold. The choice of the threshold $M_{\min}$ is up to the experimenter.

Thereby, the PCL construction by Cowan et al.

1) leads to the unavoidable overcovering (the excess probability contained in the acceptance region);

2) gives no reasonable interpretation of the value $M_{\min}$, that defines the resulting confidence belt;

3) does not solve the problem of the shrinking of the confidence intervals in the non-physical region.

**4. Feldman and Cousins recipe**

The construction of the confidence intervals suggested by Feldman and Cousins [6], as well as the method of Stern, Crow and Gardner, is based on the special order of adding the points to the acceptance region. The order is defined by the likelihood ratio.

For example, if an experiment measures the number of events *n* governed by Poisson distribution with the parameter $(\mu+b)$. Here $\mu$ is the unknown parameter that is to be estimated, *b* is the known background. Let $P(n\,|\,\mu_1)$ be the probability to obtain *n* events in the experiment with $\mu=\mu_1$. The abovementioned ratio of likelihoods is defined as follows:

$$R(n)=\frac{P(n\,|\,\mu_1)}{P(n\,|\,\mu_{best})}, \qquad (15)$$

where the value $\mu_{best}=\max(0,n-b)$ is non-negative and maximizes $P(n\,|\,\mu)$ for the given *n*. After that for each $\mu$ the points *n* are added into the acceptance region in the order of decreasing of the corresponding values of $R(n)$ until the total amount of probability in the acceptance region reaches the desired confidence level.

However, as it is noted in [6] the recipe fails to solve the problem of less then the background number of events: the confidence limit decreases with the decrease of the measured number of events. Thus one can obtain arbitrarily strong constraint on the signal regardless of the value of the background.

A similar problem occurs if one applies the recipe [6] to the bounded parameters of continuous distributions. The farther the estimator falls beyond the a priori boundary the smaller confidence interval it yields. The Feldman-Cousins recipe yields a paradoxical result: the most unreliable results (the estimators which fall far beyond the boundary) provide the strongest constraints on the estimated parameter.

The stated problems of the construction [6] lead to the situation in which it is impossible to compare not only the results of different experiments, but the results of the same experiment (e.g. two different runs) as well, if the results are presented in the form of confidence intervals constructed via Feldman and Cousins recipe.

The incomparability of the confidence intervals is intrinsic feature of the candidate solutions from the first group (see, for instance, the attempt to further modify the intervals and the order of the construction of the acceptance region in [18]).

The candidate solutions from the second group are devoid of this drawback.

## 5. Mandelkern and Shultz construction

The recipes in Sections 3 and 4 imply the use of a priori information via changing the order of construction of the acceptance region.

However these recipes lead to unphysical (short) confidence intervals near the physical boundary of the parameter. It is the result of using an estimator that does not account for the physical boundary.

Changing of the order can not change the essence of the estimation procedure: it can be done only via the choice of the estimator (its distribution contains all the information about the parameters, the experiment etc.).

Mandelkern and Shultz [8] suggested to use a modified estimator in case of bounded parameters. In the special case considered in [8] the suitable estimator is found via the method of maximum likelihood.

The procedure is as follows. The likelihood function is modified by a new factor – Heaviside function that explicitly depicts the boundary condition for the parameter. After that one obtains the estimator that always lies in the physical region for the considered study. However, the introducing of the factor seems to be somehow artificial (like a postulate). The validity of such a procedure becomes apparent only post factum, after the direct comparison of the estimate from [8] with the general solution [9].

The further construction of the confidence intervals is carried out without any additional assumptions (following the standard Neyman prescriptions). Note that the procedure is not based on Bayesian approach since the elimination of the unphysical values of the parameter is not a variant of the introducing of the uniform Bayesian a priori distribution function.

The solution [8] is as a matter of fact correct (though, once again, it can be most easily verified by direct comparison with the general solution). The Mandelkern and Shultz construction:

1) formally solves the problem of less then the background number of events (for the Poisson process with background) and the problem of shrinking of the confidence intervals for the negative values of the estimator of the nonnegative parameter of Gaussian distribution as opposed to the Feldman and Cousins recipe;

2) provides the correct amount of probability contained in the acceptance region unlike the flip-flop recipes, which violate the conditions (3) and (10).

On the other hand, the recipe of the construction of the estimator is based of the method of maximal likelihood. Thus, the construction fails to take into account a wide range of issues in which some other method of estimation is initially used.

Probably that is why the formally correct Mandelkern and Shultz approach has not been widely exploited in practice.

### 6. Method of sensitivity limit for a parameter of continuous distribution

The comprehensive solution of the issue of constructing confidence intervals for a parameter of a continuous distribution with a priori information about the limited domain of the parameter was presented in [9]. The justification of the procedure is presented below.

The defining element of the construction of confidence intervals is the estimator. What then should the estimator be instead of $\hat{\theta} = \hat{\theta}(X)$, if one knows beforehand that $\theta \geq 0$? The purpose of any estimator is to provide a value as close as possible to the unknown $\theta$. So, define a new estimator:

$$\tilde{\theta} = \max\left(\hat{\theta}, 0\right). \quad (16)$$

Evidently, $\tilde{\theta}$ yields estimates that are guaranteed to be closer to the unknown value of $\theta$ than $\hat{\theta}$, and it incorporates both the statistical information contained in the unmodified estimator $\hat{\theta}$ as well as the a priori knowledge that $\theta \geq 0$. It then remains to construct confidence intervals for the new estimator $\tilde{\theta}$.

One may wish to ponder the definition (16) prior to reading on.

The probability distribution for $\tilde{\theta}$ has the form:

$$\tilde{d}_\theta(\tilde{\theta}) = H(\tilde{\theta})\, d_\theta(\tilde{\theta}) + c_\theta \delta(\tilde{\theta}), \quad (17)$$

where $H(t)$ is the standard Heaviside step function, $\delta(t)$ is the usual Dirac $\delta$-function and

$$c_\theta = \int_{-\infty}^{0} d\hat{\theta}\; d_\theta(\hat{\theta})\,. \tag{18}$$

So, one has to deal with the aggravation of a singular contribution in Eq. **(17)**. This can be done in a regular way, or via a trick. Refer to [9] for the description of the standard approach (regularization).

Now, the trick. The key observation is as follows. The definition (16) means that the random values of the unmodified $\hat{\theta}$ are eventually carried over to the zero point and piled up there. This means that all such values will be indistinguishable: they will all yield the zero value for $\tilde{\theta}$ — and the same confidence interval. This implies that all non-positive values of $\hat{\theta}$ are to eventually yield one and the same confidence interval [0, const], where the constant is independent of $\hat{\theta}$.

Once this is understood, the construction of confidence belts can be completed entirely in terms of $\hat{\theta}$, with the only trace of $\tilde{\theta}$ being an additional condition: the resulting confidence belts must be such that all values of $\hat{\theta}$ below the a priori bound must yield the same confidence interval.

The condition has a clear experimental meaning: it can be rephrased as a requirement of robustness of confidence intervals with respect to unknown experimental (ef|de) fects. This gives the entire construction an additional physical weight. However, it is worth keeping in mind that the argument that led to it starting from the beginning of this section is transparent and specific, and does per se not need any metaphysical support: the construction in terms of the unmodified estimator $\hat{\theta}$ with the above additional condition is equivalent to a straightforward construction of confidence belts for the modified estimator (16) that incorporates the a priori information in a most straightforward and transparent fashion.

***Horizontal deformations.***

**Horizontal deformations are a visualisation of the correct modifications of confidence belts (systems of confidence intervals) that produce no undesired artefacts, e.g. an excess or lack of probability in a given confidence belt. Using the horizontal deformations one can obtain the Feldman and Cousins confidence belt as well as the confidence intervals of the method of sensitivity limit. The horizontal deformations as an instrument are to be contrasted with the so-called Flip-Flop (vertical) deformations of the confidence belts. Vertical deformations result in incorrect probability contents of the confidence belts (the conditions (3) and (10) are violated); the corresponding intervals are not confidence intervals, by definition.**

The trick of horizontal deformations is based on the following properties of allowed confidence belts for a fixed $\beta$. If, for a fixed $\theta$, $U(\theta)$ in the definition (5) is pushed down to its lower limit $U_{1-\beta}(\theta)$ then the corresponding $L(\theta) \to -\infty$ (*L* can be similarly pushed to its upper limit.) If *L* and *U* can thus be deformed while always preserving continuity and monotonicity, then there will be correctly defined inversions $u = U^{-1}, l = L^{-1}$ at every step of the deformation — i.e. an allowed confidence belt for the level $\beta$.

Given the way we draw plots in this review, such deformations occur in horizontal directions. Fig. 3 illustrates this: the fat curves must lie outside the internal belt and may only approach its boundaries (in the horizontal directions shown by the arrows) on one side at the cost of running away from it to infinity on the other side.

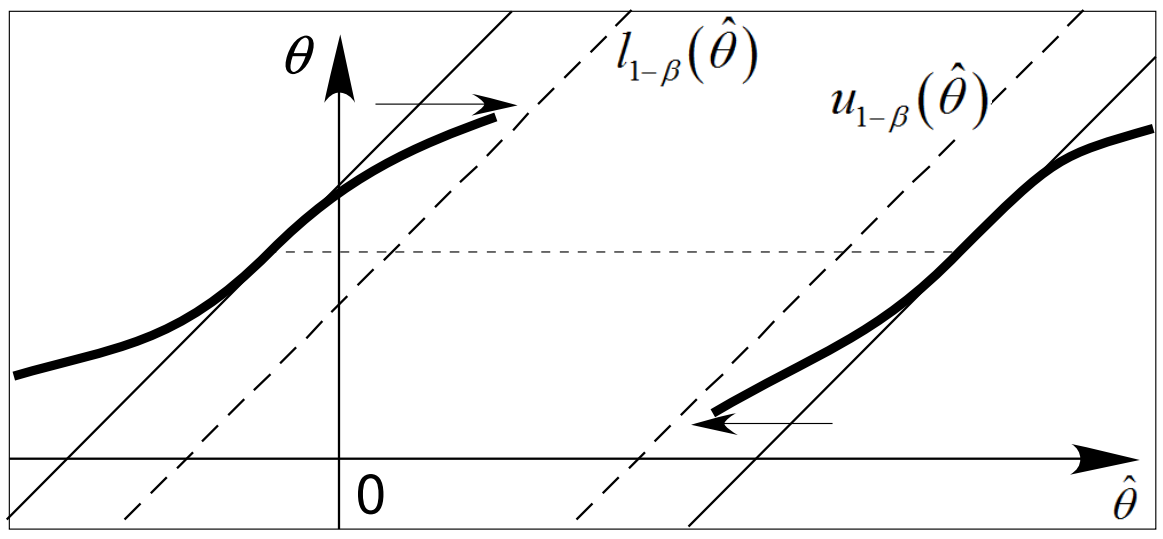

**Fig. 3.** The pairs of solid and dashed sloping curves delimit symmetric confidence belts for the confidence limits $\beta = 1-2\alpha$ and $\tilde{\beta} = 1-2(1-\beta) = 1-4\alpha$ (cf. Fig. 1). The two fat curves show a possible choice of $l, u$ for the confidence level $\beta = 1-2\alpha$. The arrows show horizontal deformations discussed in the main text.

Whenever one of the fat curves crosses a boundary of the symmetric confidence belt (the solid sloping lines) then the other fat curve crosses the other boundary, as shown with the horizontal dashed line.

The described freedom was exploited in ref. [6] where the pair *L*, *U* was chosen based on a learned belief in the magic of likelihood. The choice of [6] is illustrated in Fig. 4.

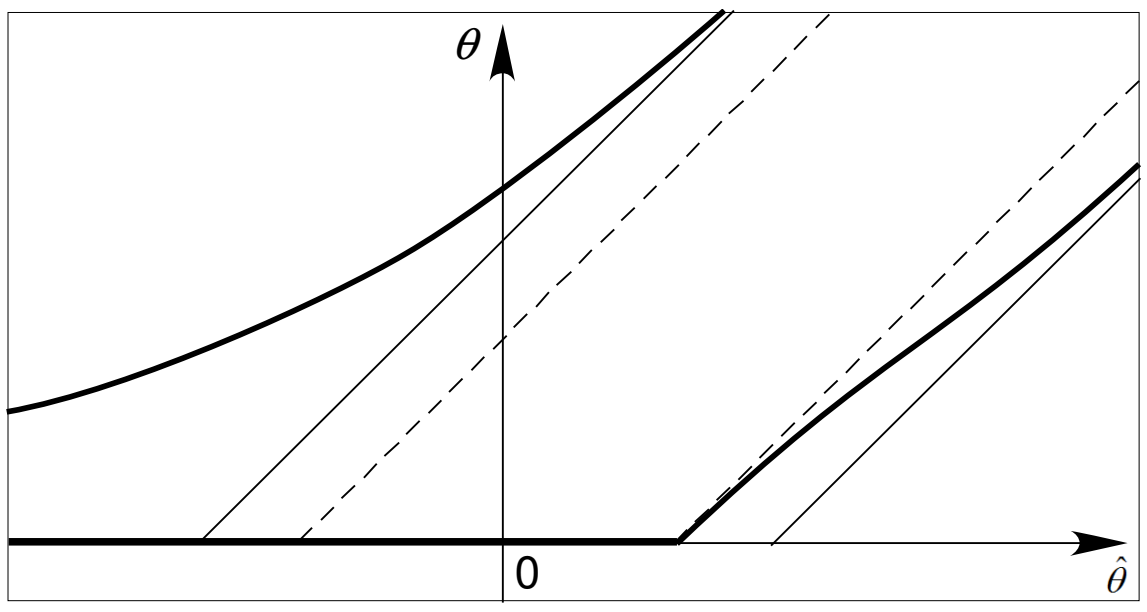

**Fig. 4.** The confidence belt (delimited by fat lines) as defined in ref. [6] (cf. their Fig. 10). The other lines are the same as in Fig. 3. The right fat curve approaching the dashed boundary

corresponds to the left fat curve running away to infinity while approaching the horizontal axis, cf. Fig. 3.

Lastly, one can take a limit deforming $L$ so that its part adheres to a part of $L_{1-\beta}(\theta)$ (see Fig. 5). This may cause $L$ (and $U$) to loose continuity at the boundary of such part. However, if the corresponding inversions $u = U^{-1}, l = L^{-1}$ continuously approach, in the limit, well-defined continuous monotonic (non-decreasing) functions of $\hat{\theta}$ , then the system of confidence intervals (6) will continuously approach a well-defined result, and the limiting confidence belt will be as good as any allowed confidence belt for the purposes of parameter estimation.

One can carry out the construction of the confidence intervals via the described trick of horizontal deformations. In the notations of Fig. 2 and with a fixed confidence level $\beta$, consider a confidence belt corresponding to two functions $l$, $u$ chosen as shown in Fig. 5 (cf. also Figs. 3 and 4).

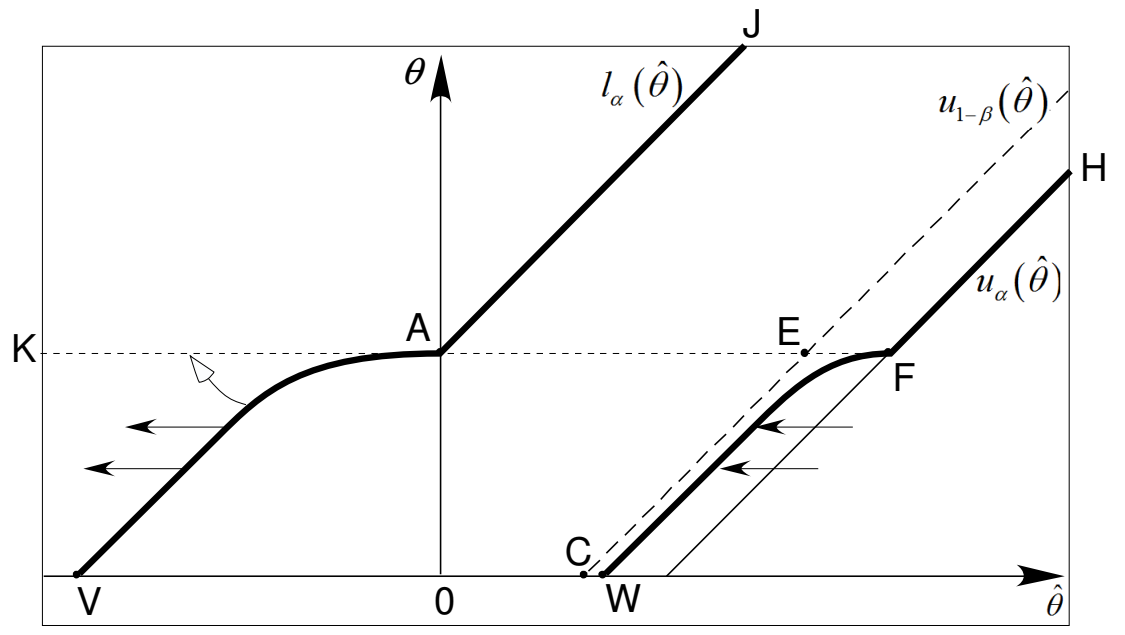

**Fig. 5.** Fat lines delimit an allowed confidence belt for $\hat{\theta}$ for the confidence level $\beta$. The other lines and points are as in Fig. 2. The fat lines are described by two functions $l$ and $u$. Black arrows indicate the horizontal deformation used to obtain the confidence belt that satisfies the additional condition.

As was discussed above, one is allowed to perform the horizontal deformation indicated by the black arrows in Fig. 5 until the curved segment WF adheres to the broken line CEF. The curved segment AV will in the end effectively be straightened out into AK (the white arrow). The limiting confidence belt is well-defined and is shown in Fig. 6.

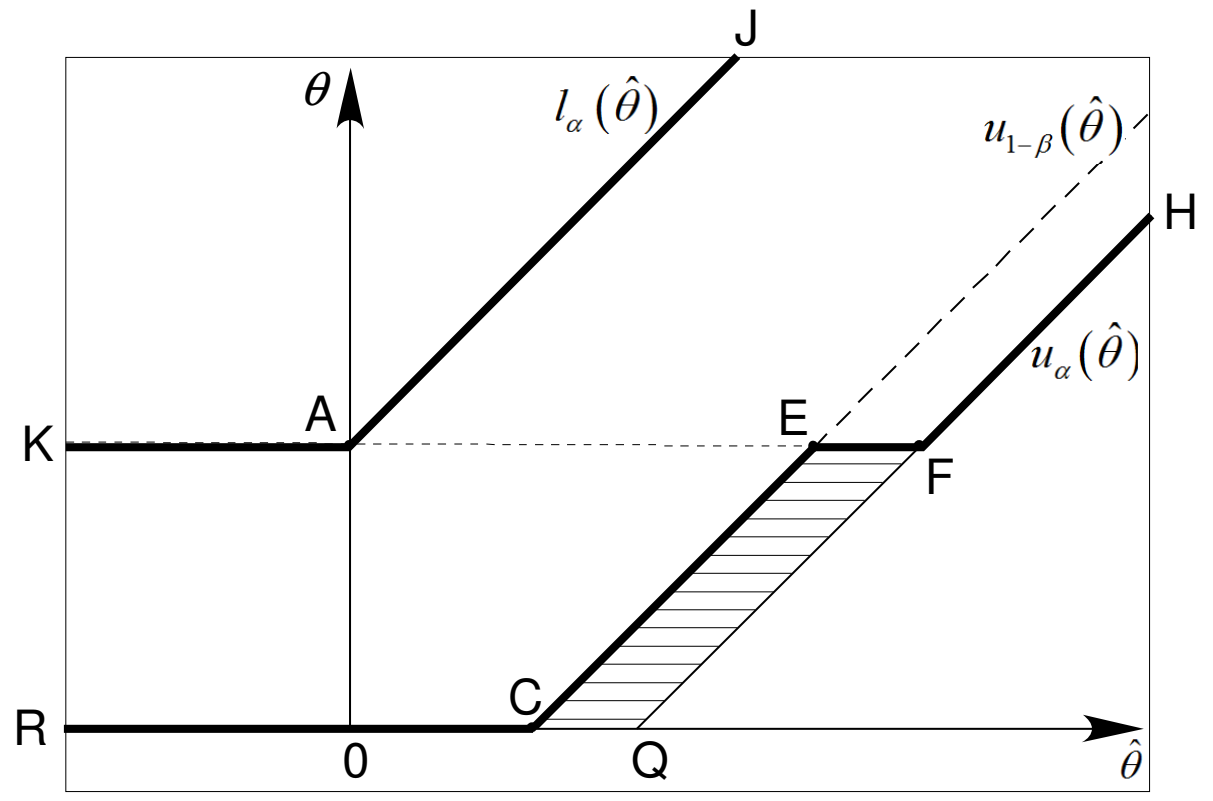

**Fig. 6.** The two fat broken lines KAJ and RCEFH delimit the resulting level-$\beta$ confidence belt for $\hat\theta$ that satisfies the additional condition and therefore correctly maps to a confidence belt for $\tilde\theta$ defined by Eq. (16). The region CEFQ cut out from the unmodified confidence belt is a pure gain obtained from the a priori knowledge.

**Confidence intervals in the method of sensitivity limit**

The resulting confidence intervals, derived within the method of sensitivity limit for the parameters of continuous distributions, are shown in Fig. 6. They can be described analytically as follows (the notations correspond to Fig. 6):

— For $\hat\theta \geq \hat\theta_\mathrm{F}$, the confidence interval $\left[u_\alpha(\hat\theta), l_\alpha(\hat\theta)\right]$ coincides with the symmetrical confidence interval for $\hat\theta$ and the confidence level $\beta = 1 - 2\alpha$, for the case of unbounded parameter.

— For $\hat\theta_\mathrm{E} \leq \hat\theta \leq \hat\theta_\mathrm{F}$, the confidence interval is $\left[\theta_\mathrm{A}, l_\alpha(\hat\theta)\right]$, i.e. the upper boundary is the same as for $\hat\theta \geq \hat\theta_\mathrm{F}$, while the lower boundary is constant and equal to $\theta_\mathrm{A}$.

— For $\hat\theta_\mathrm{C} \leq \hat\theta \leq \hat\theta_\mathrm{E}$, the confidence interval is $\left[u_{1-\beta}(\hat\theta), l_\alpha(\hat\theta)\right]$, i.e. the upper boundary is the same as above, while the lower boundary coincides with the boundary of a symmetric confidence interval for the confidence level $\tilde\beta = 1 - 2(1-\beta) = 1 - 4\alpha$.

— For $0 \leq \hat\theta \leq \hat\theta_\mathrm{C}$, the confidence interval is $\left[0, l_\alpha(\hat\theta)\right]$ with the same upper boundary and the lower boundary equal to 0.

— Finally, for all $\hat\theta \leq 0$, the confidence interval is the same, i.e. $[0, l_\alpha(0)]$.

Recall that the construction involves two symmetric confidence belts for the unmodified estimator $\hat\theta$:

1) the symmetric belt for the level $\beta = 1 - 2\alpha$, expressed in our notation as $\left[u_\alpha(\hat\theta), l_\alpha(\hat\theta)\right]$,

2) the symmetric belt for the level $\tilde{\beta} = 1 - 2(1-\beta) = 1 - 4\alpha$, expressed as

$\left[u_{1-\beta}(\hat{\theta}), l_{1-\beta}(\hat{\theta})\right]$,

and there is no need in additional cumbersome calculations.

The upper boundary of the confidence interval [0, const] that corresponds to the unphysical values of $\hat{\theta}$ resembles the so-called *sensitivity limit*. Evidently, the value of the sensitivity limit does not depend on the particular value of $\hat{\theta}$ but is defined by the uncertainty of $\hat{\theta}$. Thus the sensitivity limit represents the magnitude of the experimental error and provides an objective representation of the results of the experiment if the experimental estimate falls within the unphysical region.

## 7. Method of sensitivity limit for parameters of discrete distributions

Let us consider again the case when an experiment measures a number of events *n*, that has the Poisson distribution (9) with the parameter $\mu$. Now let us take into account the presence of background events. The number of the background events is a measured quantity, therefore, in the general case the probability distribution $P_\beta(b)$ is known. $\beta$ here stands for the unknown true value of the mean number of background events.

The study [6] considered the case when the mean number of the background events is known exactly and is equal to *b*. Then, the number of events detected during the experiment is governed by the Poisson distribution with the mean equal to $(\mu + b)$:

$$P_\mu(n) = \frac{(\mu+b)^n}{n!} e^{-(\mu+b)}. \tag{19}$$

Let us now use the additional information about the background and construct the confidence intervals for the parameter $\mu$ similar to the reasoning in [9] and Section 6. The main idea is the following: the a priori information about the parameters can be again included into the chosen estimator for the corresponding parameter. After that the constructing of confidence intervals is carried out automatically (one can compare this approach with the attempts [6] of direct modifications of the confidence intervals that contain certain arbitrariness). In the particular case when the mean value of the background is known exactly, one can use the idea from [9] and choose the quantity

$$\widetilde{(\mu+b)} = \max(n, b) \tag{20}$$

as an estimator.

Apparently, one can choose the measured value *n* as an estimator for $(\mu + b)$, but as opposed to (20), this estimator allows $\mu$ to fall below zero. Measuring *n* and using the estimator (20) after the subtraction of the constant background *b* one obtains a nonnegative estimate of the parameter $\mu$.

The probability distribution of the estimator (20) allows us to construct the confidence intervals for $(\mu + b)$ and, consequently, for the parameter $\mu$. The discreteness of the distribution (19) of *n* also allows of the following reasoning. For each given $\mu$ the probability to measure the number of event $n \le b$ is:

$$P(n \le b) = \sum_{n=0}^{[b]} \frac{(\mu + b)^n}{n!} e^{-(\mu+b)}, \tag{21}$$

where [*b*], as usual, stands for the integer part of *b*. Then, using the estimator (20) and measuring *n* one obtains the value of the estimator equal to *b* with the probability (21).

Hence, the probability distribution for the variable $\max(n,b)$ consists of the distribution (19) for the values $n > b$ and the probability (21) to obtain the value *b* during the measurements (Fig. 7).

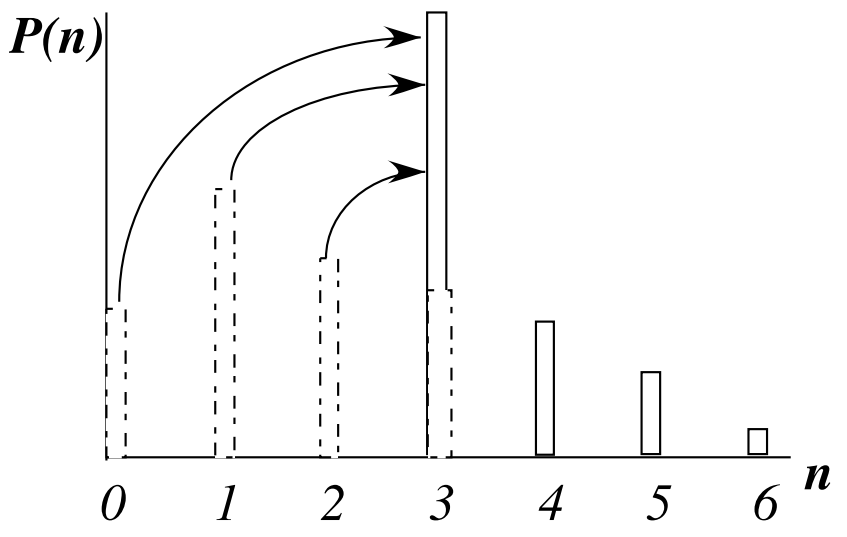

**Fig. 7.** Modification of the probability distribution due to the transition form the conventional estimator $\widetilde{(\mu + b)} = n$ (dashed plot in the range $n \le b$) to the modified estimator $\widetilde{(\mu + b)} = \max(n,b)$ (solid plot) when the mean value of the background is equal to $b = 3$. In the range $n > b$ the probabilities for both conventional and modified estimators coincide.

Using the probability distribution and following the procedure from Section 2(b), one can construct the confidence intervals for the quantity $(\mu + b)$ and, thus, for the parameter $\mu$.

It can be more convenient for the practical use to present the confidence intervals in terms of variables $\mu$ and *n* (Fig. 8). In this form one can immediately obtain the confidence interval for the parameter $\mu$ for each measured value $n_0$.

Note that all the values $n \le b$ become indistinguishable after the transition to the estimator (20). Therefore, the same confidence interval corresponds to all these values of *n*. The upper bound of such an interval corresponds to so-called sensitivity limit. It contains the information on the magnitude of background in the given experiment.

**Confidence intervals for a parameter of discrete distribution in the method of sensitivity limit.**

The confidence belt for the parameter of Poisson distribution with known background constructed via the method of sensitivity limit is shown in Fig. 8 (confidence level is $\beta = 90\%$ ). The belt can be described analytically as follows:

— for all $n > b$, the upper boundary of the belt coincides with the boundary of a symmetric confidence interval for the confidence level $\beta = 90\%$ ;

— for $n \leq b$, the upper boundary is constant and equal to $\mu_b$ .

The lower boundary is divided into four parts (similarly to the continuous case):

— for small $n$, the lower boundary is 0;

— the next part of the lower boundary coincides with the boundary of the one-side interval for the confidence level $\beta = 90\%$ ;

— the transitional part between the boundaries of the one-sided and the symmetrical interval: the lower boundary is equal to $\mu_b$ ;

— for large $n$, the lower boundary coincides with that of the symmetric confidence interval for the confidence level $\beta = 90\%$ .

Confidence intervals within the method of the sensitivity limit for the Poisson process with background can be calculated with a dedicated software [20], since there are no simple analytical formulae here (see the implicit formula (13); the software uses a search algorithm to determine $\mu^*$ ).

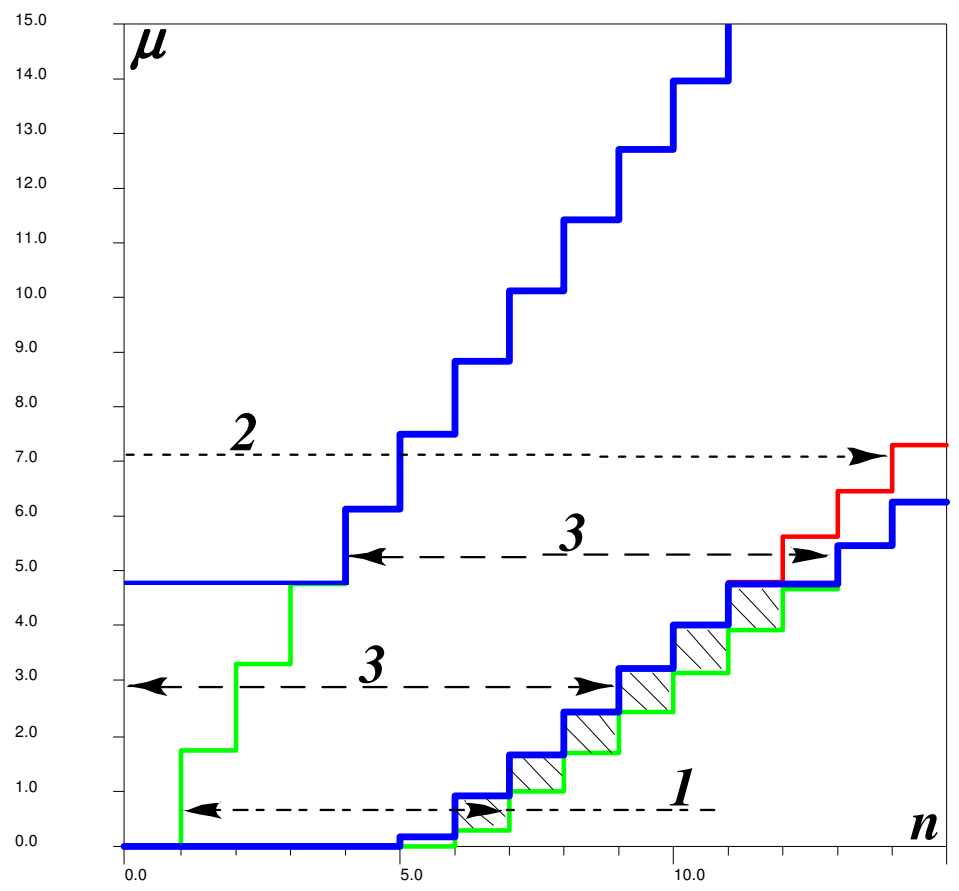

**Fig. 8.** Confidence intervals for the unknown signal $\mu$ with the Poisson background (the mean value $b=3$): green (1) – 90% C.L. symmetric confidence belt with no account of the information about the background; red (2) – one-sided 90% C.L. confidence interval with no account of the information about the background; blue (3) – 90% C.L. confidence belt, based on the estimator (20) of the parameter $\mu$ with account of the a priori information.

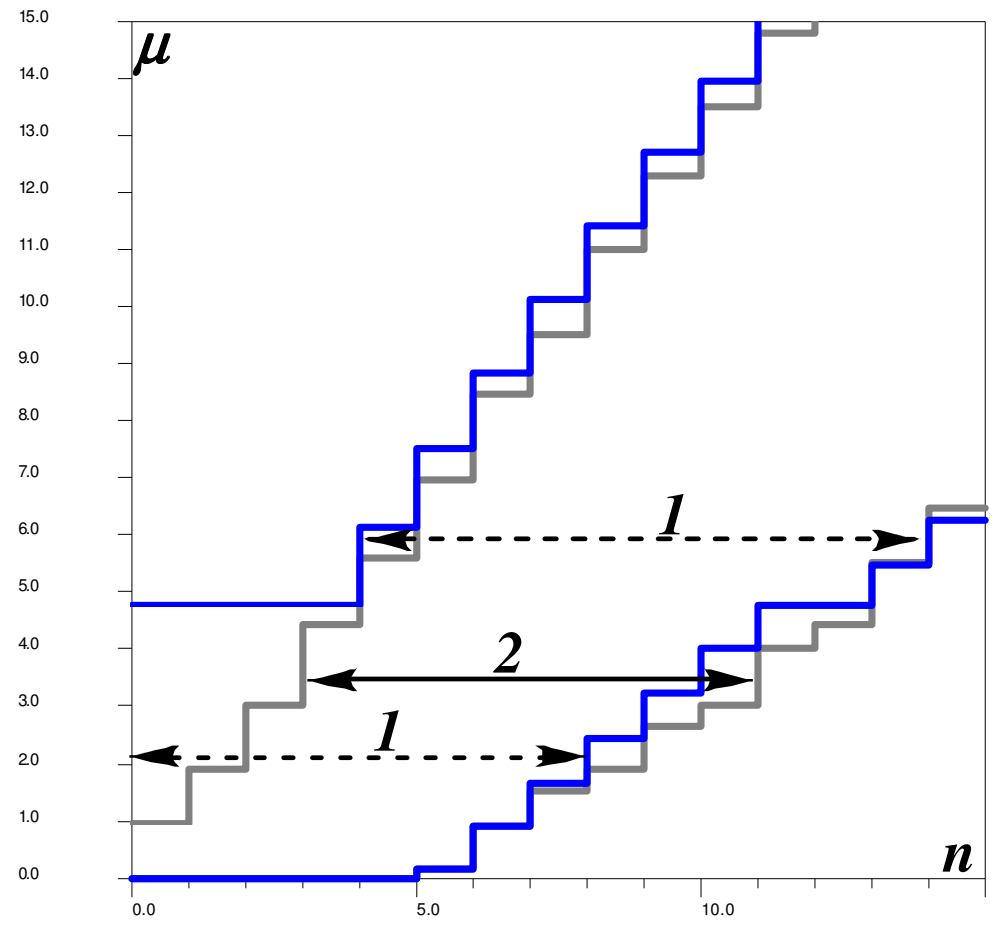

**Fig. 9.** Confidence belts for the unknown signal $\mu$ with the Poisson background (the mean value $b=3$): blue (1) – 90% C.L. confidence belt with account of the a priori information, similar to Fig. 8., grey (2) – 90% C.L. confidence belt, constructed via Feldman and Cousins recipe [6].

Fig. 9 presents the confidence belt constructed via Feldman and Cousins recipe (grey) and the one constructed with the use of the a priori information in the estimator (blue). It is apparent that the confidence belt with the correct use of the a priori information (blue plot in Fig. 9)

1) provides the correct estimate of the parameter in the region $n \le b$;

2) has an area analogous to the area CEF in Fig. 6; therefore it provides the best possible estimate of the lower bound of the confidence interval;

3) guarantees by construction that the amount of probability in the acceptance region is close to 90% (it is impossible to obtain exact 90% amount of probability in the case of discrete distributions).

One can also use Table 1 (in Appendix) to compare the construction with the correct usage of the a priori information with the construction [6]. Table 1 is similar to Tables II-IX in [6] and it provides the confidence intervals for the 90% C.L. and various values of background and measured number of events ($b$=(0..10), $n_0 = (0..20)$).

The software for the calculation of the confidence intervals for various confidence levels and parameters $b$ and $n_0$, and for the drawing of the corresponding confidence belts is available for downloading at the URL [20].

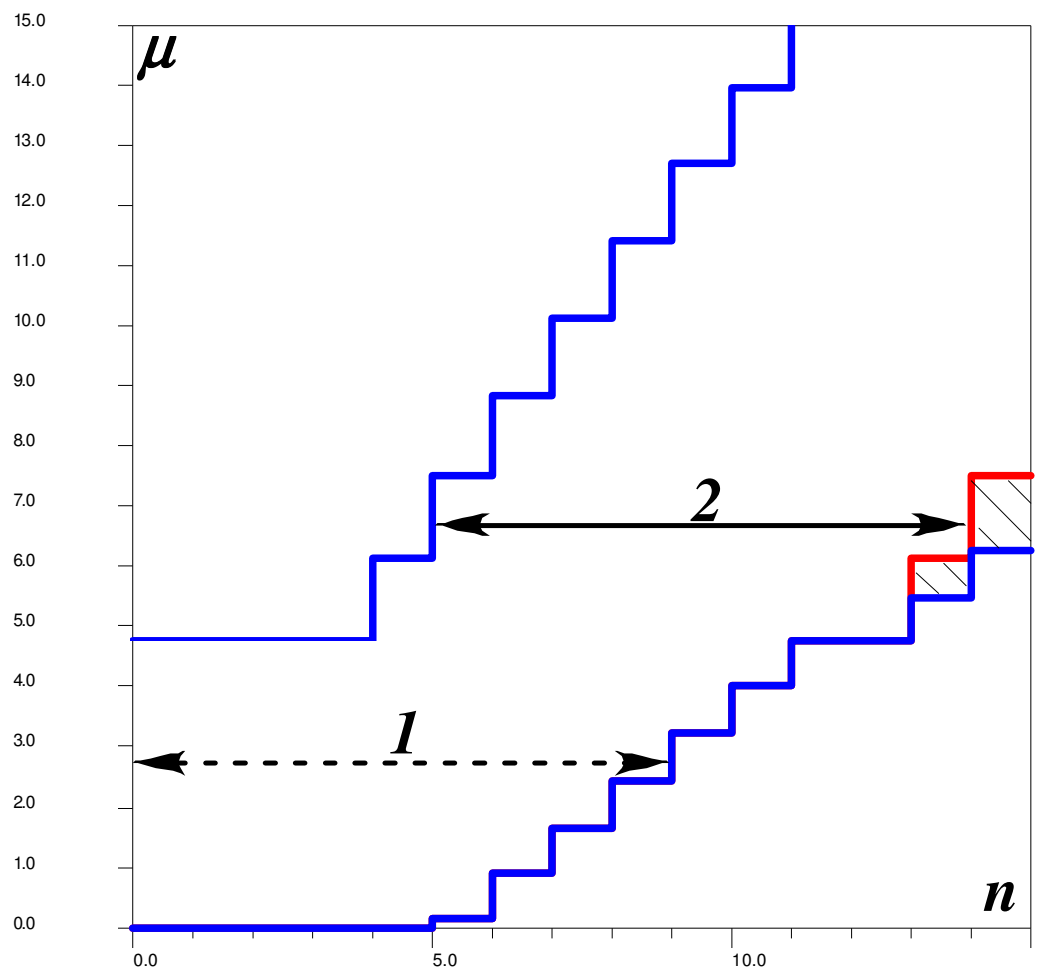

**Fig. 10.** Confidence belts for the unknown signal $\mu$ with the Poisson background (the mean value *b=3*): blue (1) – 90% C.L. confidence belt with account of the a priori information, red (2) – 90% C.L. confidence belt with account of the a priori information, constructed for the special case when the upper boundary of the confidence belt for the parameter $\mu$ is of the most importance.

Due to the additional freedom in construction of confidence intervals for parameters of discrete distributions (the transition from the exact equality (11) to the enequality ≥ in Eq. (10)), one can minimize the amount of probability contained in the acceptance region and draw it near to the required confidence level (in our examples – 90 %). For instance, the region to the right of the upper boundary in Fig. 10 satisfy the condition $P_{\tilde{\mu}}\left(n_1(\mu) \geq n\right) \geq 0.95$. This upper boundary represents the one-sided 95% confidence interval. As it was mentioned in Section 2(b), the lower boundary can be constructed formally according to the condition $P_{\tilde{\mu}}\left(n_2(\mu) \leq n\right) \geq 0.95$ (blue plot in Fig. 10). That immediately leads to the condition $P_{\tilde{\mu}}\left(n_1(\mu) \geq n \geq n_2(\mu)\right) \geq 0.90$ to be satisfied.

On the other hand, if the upper boundary of the confidence belt is fixed one can construct the lower boundary $n_2(\mu)$ directly from the condition $P_{\tilde{\mu}}\left(n_1(\mu) \geq n \geq n_2(\mu)\right) \geq 0.90$ (red plot in Fig. 10.) The amount of probability in the acceptance region within the red belt is less or equal to the amount of probability within the blue confidence belt. Thus, one can obtain the confidence belts that are closer to the required confidence level by rejecting the symmetry of the resulting confidence belts.

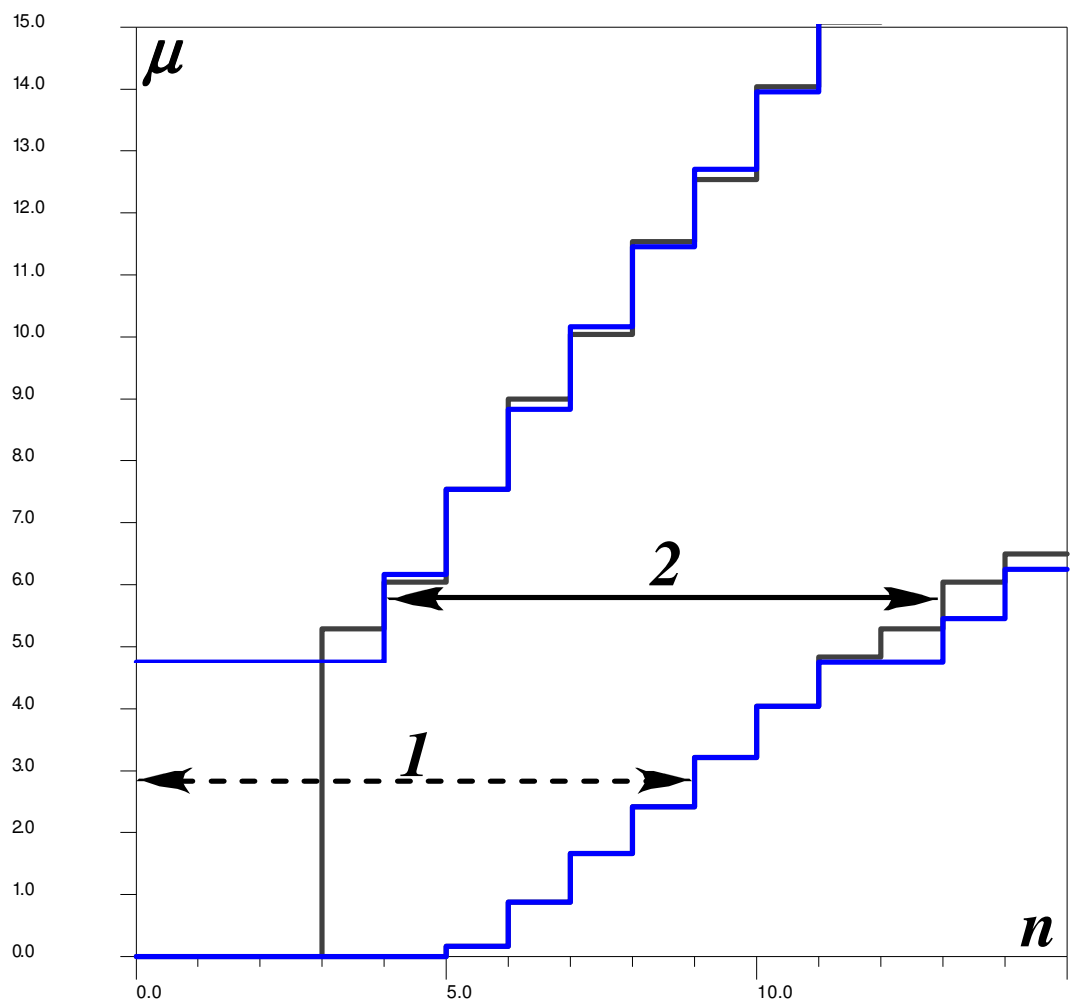

**Fig. 11.** Confidence belts for the unknown signal $\mu$ with the Poisson background (the mean value *b=3*): blue (1) – 90% C.L. confidence belt with account of the a priori information, grey (2) – 90% C.L. confidence belt for the estimator (20), constructed via Sterne, Crow and Gardner procedure (the shortest set of confidence intervals).

For completeness sake, one can also compare (Fig. 11) the confidence belt, constructed with the correct account of the a priori information (blue plot), with the interval, constructed via Sterne, Crow and Gardner procedure (grey plot). The Sterne, Crow and Gardner construction was carried out for the estimator (20). The grey plot yields an empty confidence interval in the range $n < b = 3$ because the estimator (20) has no values in that region. This set of confidence intervals has by construction the smallest length for the given confidence level.

As a result of introducing the a priori information into the estimator one obtains the set of confidence intervals with the following remarkable properties:

— the problem of less then the background measured number of events pointed out in [6] is solved (in the region where the number of events is less then the background the confidence belt immediately yields the upper bound for the estimated parameter $\mu$, this estimate does not depend on the value of $n$, as in the region $n \leq 3$ in Fig. 8);

— the lower boundary of the confidence belt has the area analogous to the area CEF in Fig. 6;

— due to the ambiguity of the definition of the confidence intervals for the discrete distributions (condition (10)) there are various choices of construction of the confidence belts (with fixed upper or lower boundaries, Sterne, Crow and Gardner construction and their combinations). The choice depends on the particular situation, for instance, on the necessity to get more reliable upper or lower limits for the parameters.

Therefore, the correct account of the a priori information about the background via the choice of a suitable estimator provides the confidence intervals that meet the requirements of physical reliability and devoid of the flaws of the constructions [6], [18] and [7]. Similarly to the case of the continuous distributions, the considered approach allows one to compare the confidence intervals obtained in different experiments.

## 8. The best upper limit within the method of sensitivity limit

### a) Continuous distributions

Let us consider now the case of a non-symmetric confidence interval. The case can be important for various experimental applications.

In Section 6 a conventional estimator $\hat{\theta}$ for the parameter $\theta$ is redefined in the way to take into account the a priori inequality $\theta \geq 0$. Then for the redefined estimator $\tilde{\theta} = \max\left(\hat{\theta}, 0\right)$ (Eq. (16)), a conventional confidence belt was constructed in a more or less straightforward fashion. The treatment of the $\delta$-functional contribution to the probability distribution of $\tilde{\theta}$ was simplified via an observation that reduced the problem to constructing a confidence belt for the unmodified estimator $\hat{\theta}$ in such a way that the resulting belt satisfies an additional condition (see Section 6). The construction was accomplished using the trick of so-called horizontal deformations, with the result represented by Fig. 6.

Section 6 modified the standard symmetric confidence belt, which corresponds to the option $\alpha = \alpha' = (1-\beta)/2$ in terms of Fig. 1. A natural variation on the same theme is to accomplish a similar modification for the asymmetric case $\alpha' = 0,\ \beta = 1-\alpha$, that corresponds to an upper bound for the confidence level $\beta$:

$$\mathsf{P}\left(\theta < l_{1-\beta}(\hat{\theta})\right) = \beta . \tag{22}$$

This option is useful when one is trying to measure a positive signal whereas the statistical accuracy may not be high enough to establish a non-zero signal with a high confidence. Then one would like to establish as tight an upper bound as possible. This problem definition and the corresponding solution were considered in [12].

In [12] the confidence belt (22) is modified to accommodate the a priori inequality $\theta \geq 0$.

The required geometrical infrastructure is provided by Fig. 12 that differs from Fig. 2 by adding a few more intersection points (the intersection points MBDN on the horizontal line LG).

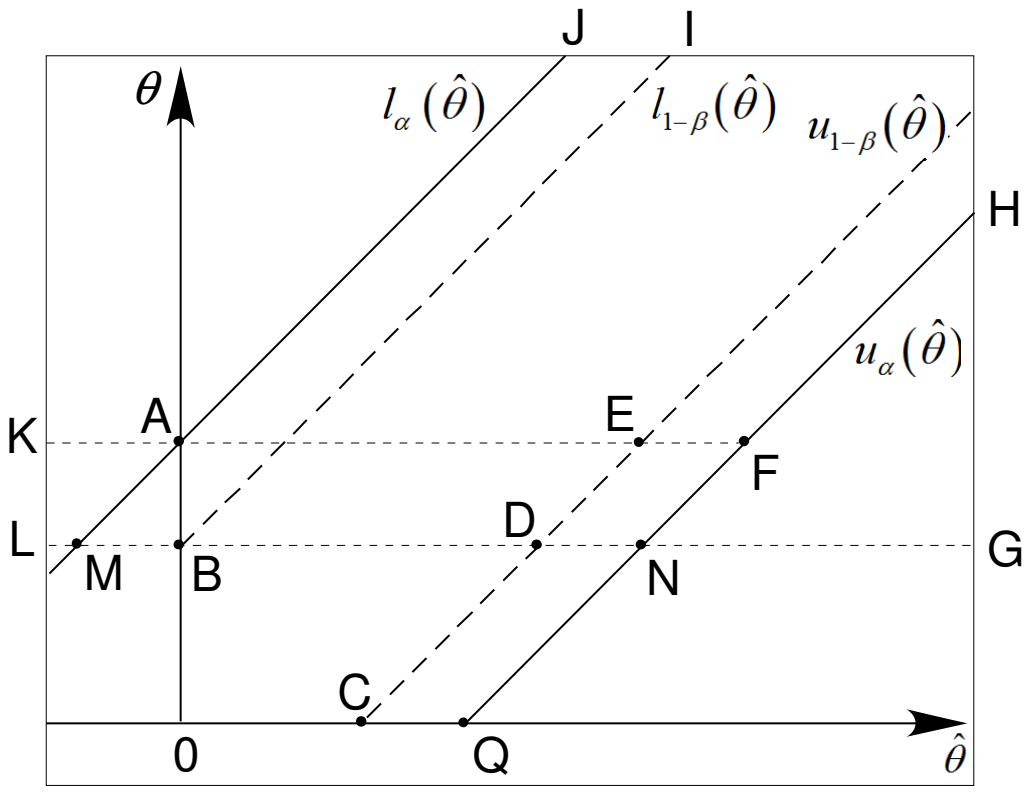

**Fig. 12.** The pairs of solid and dashed sloping lines delimit symmetric confidence belts for the confidence limits $\beta = 1-2\alpha$ and $\tilde\beta = 1-2(1-\beta) = 1-4\alpha$. The functions that correspond to the lines are shown in the figure. A and B are intersections with the vertical axis of the lines $\theta = l_\alpha(\hat\theta)$ and $\theta = l_{1-\beta}(\hat\theta)$. Points A and B determine the horizontal lines KF and LG along with further intersection points.

Only the points M, B, C, D, N will play a role in what follows; the other points are shown to establish a connection with Fig. 2.

The number $\theta_B$ is the vertical position of the intersection point B (and of M, D, and N):

$$\theta_B = l_{1-\beta}(0)\,. \tag{23}$$

The numbers $\theta_C < \theta_D$ are the horizontal positions of the points C and D:

$$\theta_C = U_{1-\beta}(0), \quad \theta_D = U_{1-\beta}(\theta_B)\,. \tag{24}$$

The unmodified bound (22) corresponds to confidence intervals (level $\beta$) that start on the upper dashed line BI and stretch down to infinity.

To obtain a modified version of the bound (22), one starts from an allowed confidence belt $\left[u(\hat\theta), l(\hat\theta)\right]$, shown with the fat lines in Fig. 13.

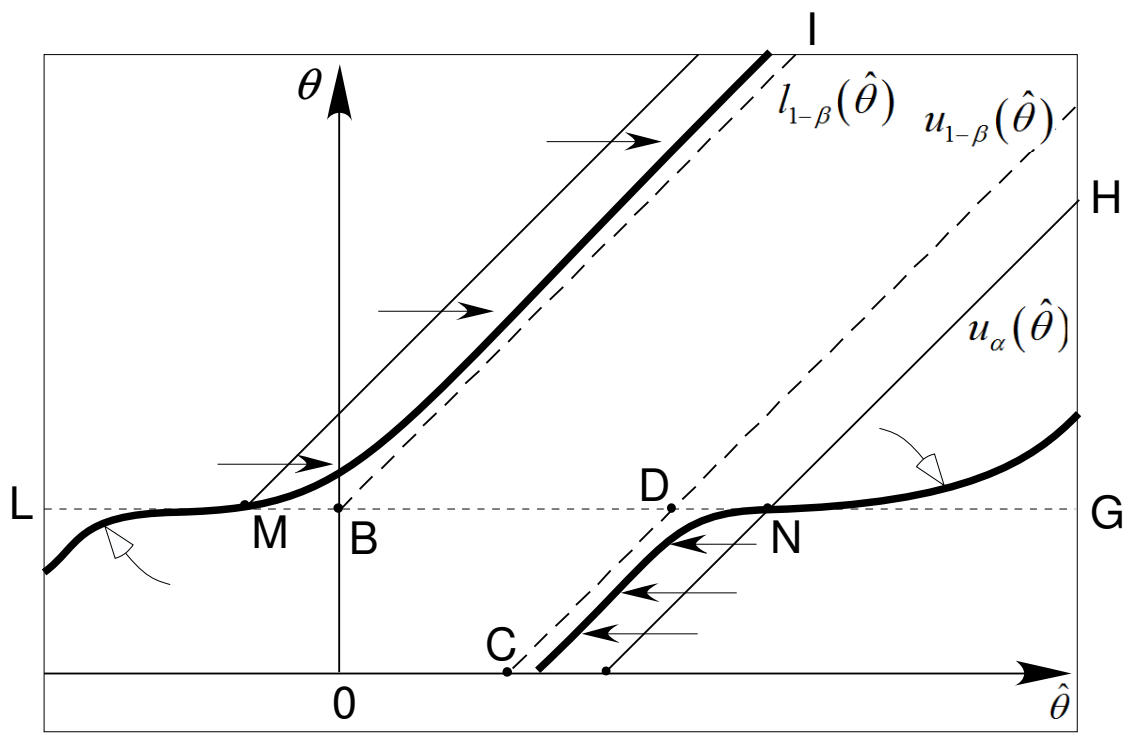

**Fig. 13.** The two fat curves delimit an allowed confidence belt for the confidence level $\beta$. The fat lines are hinged at the points M and N. Black arrows show allowed horizontal deformations. White arrows show the resulting straightening of the corresponding segments.

Then one performs the horizontal deformations of $u$ and $l$ as shown by the black arrows (detailed explanations of the trick are given in Section 6). Then the lower segment of $u$ below point N is pressed to the straight segment CD, whereas the upper segment of $l$ above point M is pressed to BI. The resulting effective deformations on the other side are shown by white arrows. The confidence belt thus obtained is shown in Fig. 14.

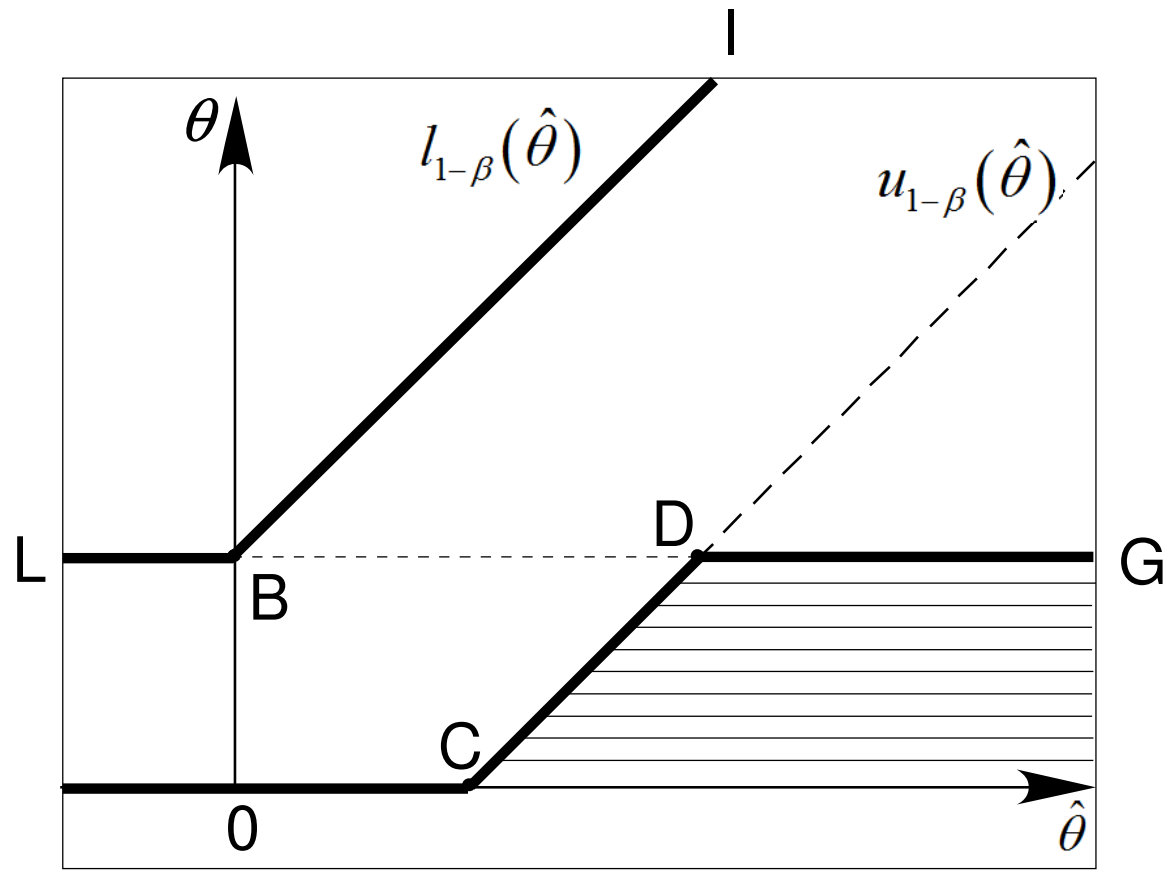

**Fig. 14.** The confidence belt obtained from the unmodified upper bound for the confidence level $\beta$ by taking into account the a priori information $\theta \geq 0$. The region under CDG is a pure gain from the a priori information.

**Non-symmetric confidence belts in the method of sensitivity limit**

The confidence belt for the best upper limit, modified via the method of sensitivity limit is shown in Fig. 14. The analytical description is as follows (we are talking about the confidence level $\beta$; the positions of points B, C and D along the $\hat{\theta}$ axis in Fig. 14 are denoted as $\hat{\theta}_B, \hat{\theta}_C$ and $\hat{\theta}_D$, respectively):

— For $\hat{\theta} \geq \hat{\theta}_D$, the confidence interval is $\left[\theta_B, l_{1-\beta}(\hat{\theta})\right]$, i.e. the upper bound is the same as in the unmodified case, eq. (22), but restricted from below at $\theta_B$. The region under CDG is exactly the gain from the a priori information.

— For $\hat{\theta}_C \leq \hat{\theta} \leq \hat{\theta}_D$, the confidence interval is $\left[u_{1-\beta}(\hat{\theta}), l_{1-\beta}(\hat{\theta})\right]$, i.e. exactly the unmodified symmetric confidence interval for the confidence level $\tilde{\beta} = 1 - 2(1-\beta) = 1 - 4\alpha$.

— For $0 \le \hat{\theta} \le \hat{\theta}_{\mathrm{C}}$, the confidence interval is $\left[0, l_{1-\beta}(\hat{\theta})\right]$, i.e. the unmodified bound (22) restricted from below by the physical boundary.

— Lastly, for all $\hat{\theta} \le 0$ the confidence interval is fixed as $[0, \theta_{\mathrm{B}}]$.

The noteworthy properties of this confidence belt are as follows:

— The estimate is robust for non-physical values of the estimator, i.e. for $\hat{\theta} < 0$.

— The interval's upper bound for physical values of $\hat{\theta}$ is the same as in the unmodified case (22) and is the lowest possible one at the confidence level $\beta$.

— The interval's lower bound breaks off zero at the earliest point possible for the given confidence level ($\theta_{\mathrm{C}}$), and the lower bound is maximal possible for this confidence level in the interval $\hat{\theta}_{\mathrm{C}} \le \hat{\theta} \le \hat{\theta}_{\mathrm{D}}$.

— Neither complicated algorithms nor tables are required on top of the standard routines to compute the confidence interval for the confidence level $\tilde{\beta} = 1 - 2(1-\beta) = 1 - 4\alpha$.

— One more important feature of the upper limit is that it is devoid of the overcoverage (exceeding of the confidence level) that is inherent in some artificial recipes [7].

**б) Discrete distributions**

The construction of the upper limit with a priori information is carried out similarly to that in the case of a continuous distribution.

First, one chooses the estimator in the form of (20). Then, much as in Fig. 12 and 13, one considers the confidence intervals: 90% C.L. two-sided with no account of the a priori information and the upper and lower boundaries of 90% C.L. one-sided confidence intervals (Fig. 15). Although the procedure of horizontal deformation is not defined for the discrete distributions, with the use of the estimator (20) one obtains the acceptance region (bounded by black lines in Fig. 15) similar by the structure to the confidence belt in Fig. 14.

**The method of sensitivity limit for the parameter of Poisson distribution with background: a non-symmetric confidence belt**

The non-symmetric confidence belt for the best upper limit (the confidence level is $\beta = 90\%$) constructed via the method of sensitivity limit is shown in Fig. 15. It can be described analytically as follows.

The upper boundary consists of two parts:

— for $n \le b$, the upper boundary is constant and equal to $\mu_b$;

— for $n > b$, the upper boundary coincides with the boundary of the one-sided interval with confidence level $\beta = 90\%$.

The lower boundary consists of three parts:

— for small $n$, it is equal to 0;

— the second part coincides with the lower boundary of the one-sided interval with confidence level $\beta = 90\%$;

— finally, when the lower boundary reaches the value $\mu_b$, it becomes constant.

Similarly to the symmetric intervals for the Poisson process with background (see Section 7) it is convenient to do the calculations with the dedicated software [20].

The best upper limit for the unknown signal $\mu$ with the Poisson background with the mean value $b$ for the estimator (20) possesses the following properties.

— Provided the measured number of events is $n \leq b$, the construction yields the same confidence interval. That solves the problem of less than the background number of measured events pointed out in [6].

— The amount of probability contained in the acceptance region is close to the required confidence level. The exceeding of the confidence level (overcoverage in terms of [6]) is an undesirable feature of some other constructions [19].

— The lower boundary of the confidence interval moves away from the axis $n$ (becomes non-zero) at the lowest possible value of $n$. Thus, not only the best (the most stringent) upper limit is obtained, but one can also claim the detection of signal at rather small values of the measured number of events (the non-zero lower boundary of a confidence interval can be somehow interpreted as presence (or a signature) of the signal).

— As in the case of continuous distributions the area beneath the lower boundary is a gain from using the a priori information about the background.

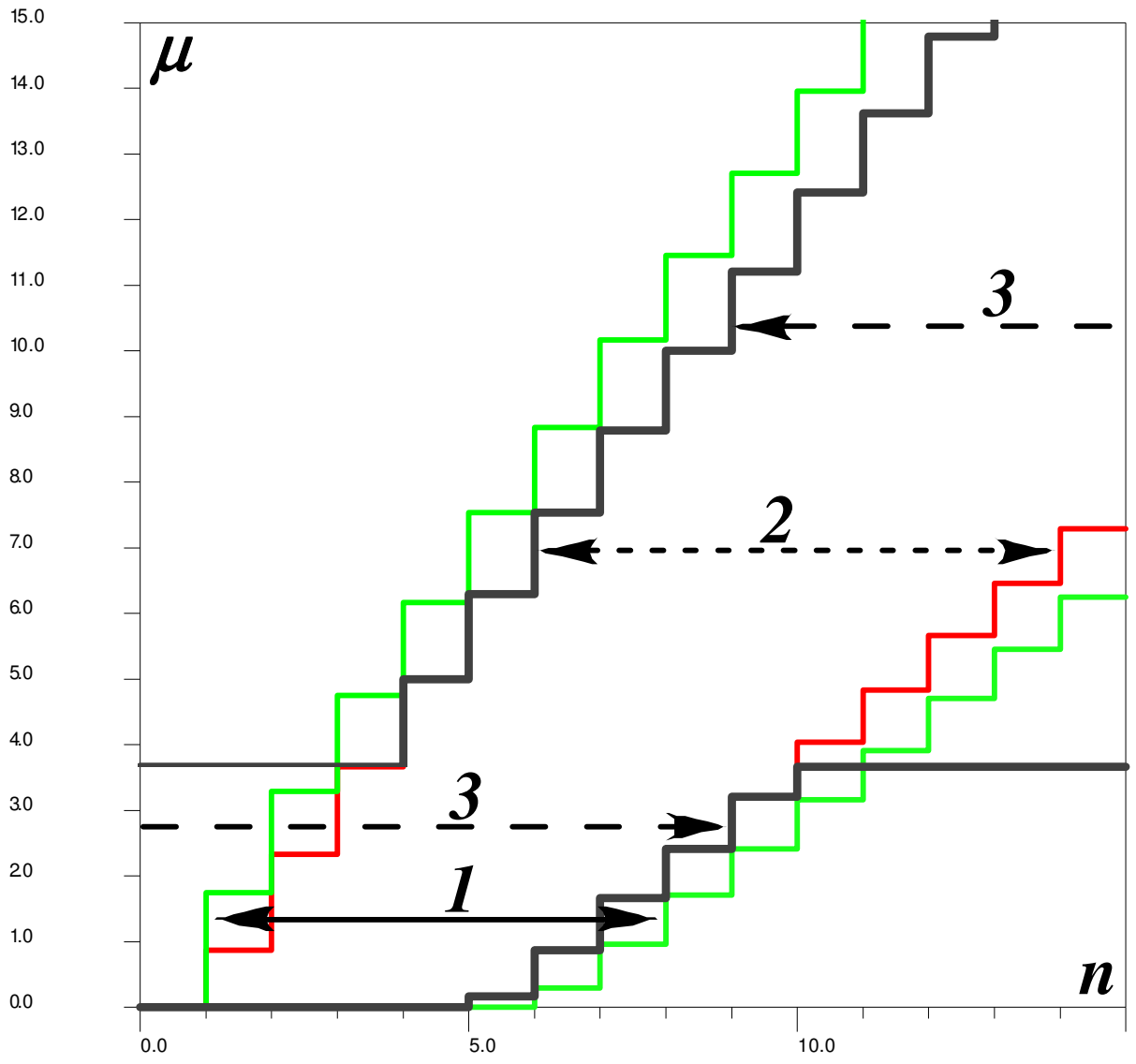

**Fig. 15.** The upper boundary of the confidence interval for the unknown signal $\mu$ with the Poisson background (the mean value $b=3$): green (1) – 90% C.L. symmetrical confidence interval with no a priori information taken into account, red (2) – 90% C.L. upper and lower one-sided intervals with no a priori information, black (3) – 90% C.L. one-sided confidence interval for the estimator (20) – the best upper limit similarly to Fig. 14.

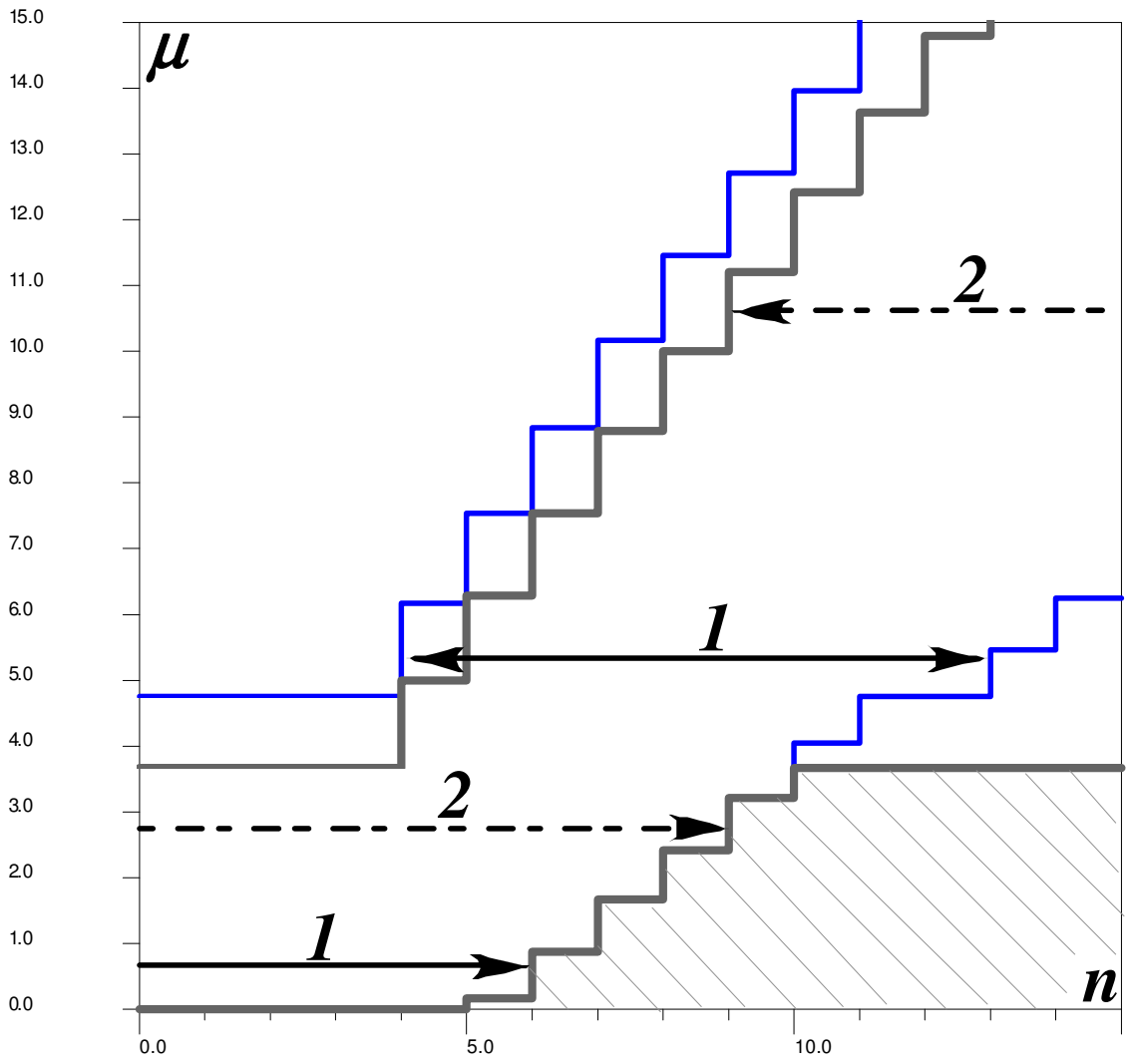

**Fig. 16.** Confidence intervals for the unknown signal $\mu$ with Poisson background (the mean value $b=3$): blue (1) – 90% C.L. two-sided confidence interval with a priori information, grey (2) – 90% C.L. confidence interval for the estimator (20) – the best upper limit similarly to Fig. 14.

For the sake of comparison, let us also consider the two-sided (symmetric) 90% C.L. confidence belt with the a priori information (blue plot in Fig. 16) and the 90% C.L. best upper limit (grey plot in Fig. 16) for the estimator (20). One can easily see that the best upper limit for the parameter $\mu$ falls below the upper boundary of the symmetric confidence belt. The lower boundaries of both belts coincide for all values of *n* up to $n = 10$. It is apparent that the symmetric interval provides a universal estimate for the parameter – for any measured number of events. The confidence belt for the best upper limit also exists (by construction) for any measured value *n*. However, the most interesting application for this confidence belt lies in small values of measured events *n*, (in our examples, $n < 10$). In this particular region the upper limit is the most stringent while the detection of the signal (in the abovementioned sense, i.e. a non-zero value of the lower boundary) is still possible.

We emphasize once again that the correct construction of the upper limit with the use of a priori information for continuous and discrete distributions allows one to directly compare the experimental results presented in form of confidence intervals.

## 9. Mass estimates of the electron antineutrino in the Mainz and Troitsk experiments

The properties of the confidence intervals constructed via the method of sensitivity limit can be illustrated with the help of the results of the two experiments on neutrino mass measurements in tritium $\beta$-decay in Mainz and Troitsk. Both experiments measure the squared mass of electron antineutrino. The parameter is nonnegative, but its estimate can fluctuate into the negative values. Therefore a correct procedure for constructing the upper limit for the neutrino mass is required.

The Mainz experiment [10] yielded the estimates for the neutrino mass squared $\tilde{m}_\nu^2 = -1.2\ eV^2$ and $\tilde{m}_\nu^2 = -0.6\ eV^2$ (depending on the choice of one of the parameters of the experimental setup, namely the excitation probability of neighbour molecules in the tritium source). Apparently, the negative values of $\tilde{m}_\nu^2$ appear due to underestimation of systematic factors or due to statistical fluctuation. The combined uncertainty is $\Delta m_\nu^2 = 3.04\ eV^2$. With these values one can obtain the upper limits on the neutrino mass via the Feldman and Cousins recipe, for instance, at 95% confidence level. For this one finds in the Table X in [6] the upper boundaries of confidence intervals for $\tilde{m}_\nu^2 = -0.4 \cdot \Delta m_\nu^2$ and $\tilde{m}_\nu^2 = -0.2 \cdot \Delta m_\nu^2$: 1.77 and 1.58, correspondingly (all the values are given in the units of the uncertainty). Then, multiplying these two values by the value of the uncertainty and extracting the square root, one obtains the following estimates for the neutrino mass: $m_\nu < 2.2\ eV$ and $m_\nu < 2.3\ eV$.

However, it is asserted in [10] that the estimate $\tilde{m}_\nu^2 = -0.6\ eV^2$ was obtained in a consistent analysis and is more reliable than the value $\tilde{m}_\nu^2 = -1.2\ eV^2$. Therefore, the recipe [6] provides a paradoxical result: it yields a better constraint ( $2.2\ eV$ ) for more negative and less reliable estimate $\tilde{m}_\nu^2$.

Hence, the recipe [6] requires additional information about the estimate. Moreover, the specific value of the estimate from the non-physical region has a considerable influence on the value of the upper limit. Thus, indicating only two values $2.2\ eV$ and $2.3\ eV$ as constraints on the neutrino mass is not enough for choosing the best of the estimates. Recall that the very aim of constructing confidence intervals is to indicate a number that can be directly used for comparing various experiments (in our example – various methods of data analysis of the same data), with no additional information on the experimental estimates.

Let us now apply the method of sensitivity limit to the results of the Mainz experiment. For the method of sensitivity limit the estimates from the non-physical region $\tilde{m}_\nu^2 < 0$ are indistinguishable, and the upper boundary of the confidence interval depends only on the value of uncertainty (see Fig. 6 and Fig. 14). At the 95% confidence level the upper limit is $1.96 \cdot \Delta m_\nu^2$. For the Mainz experiment one obtains

$$m_\nu < 2.4\ eV = \sqrt{1.96 \cdot 3.04\ eV^2}$$

for both estimates of the neutrino mass squared $\tilde{m}_\nu^2 = -1.2\ eV^2$ and $\tilde{m}_\nu^2 = -0.6\ eV^2$. No additional information about the fitting method or other parameters is required. Note that when the estimate for the neutrino mass squared is positive, the confidence interval can be also obtained directly from the confidence belt constructed via the method of sensitivity limit (see Fig. 14 for $\tilde{\theta} > 0$).

To show how handy is the method of sensitivity limit in comparing the results of different experiments, consider the results of the Troitsk-nu-mass experiment [2]. The estimate for the neutrino mass squared in [2] is $\tilde{m}_\nu^2 = -0.67\ eV^2$, while the uncertainty is $\Delta m_\nu^2 = 2.53\ eV^2$. Since the estimate again falls into the non-physical region, the upper boundary of the confidence interval depends only on the value of the uncertainty and is equal to (see Fig. 17):

$$m_\nu < 2.2\ eV = \sqrt{1.96 \cdot 2.53\ eV^2} = \sqrt{4.96\ eV^2}.$$

From the two values, $2.4\ eV$ (Mainz) and $2.2\ eV$ (Troitsk), both obtained via the method of sensitivity limit, one can immediately infer that the experiment in Troitsk yielded a better constraint on the mass of the electron neutrino. One need not refer to the issue of the negative estimates of the neutrino mass squared and there is no need in additional information on the fitting procedure and other fitting parameters. The possibility to directly compare the results of different experiments is the major advantage of the method of sensitivity limit. With the Feldman and Cousins recipe such a comparison is incorrect, furthermore it can result in a false interpretation of experimental results.

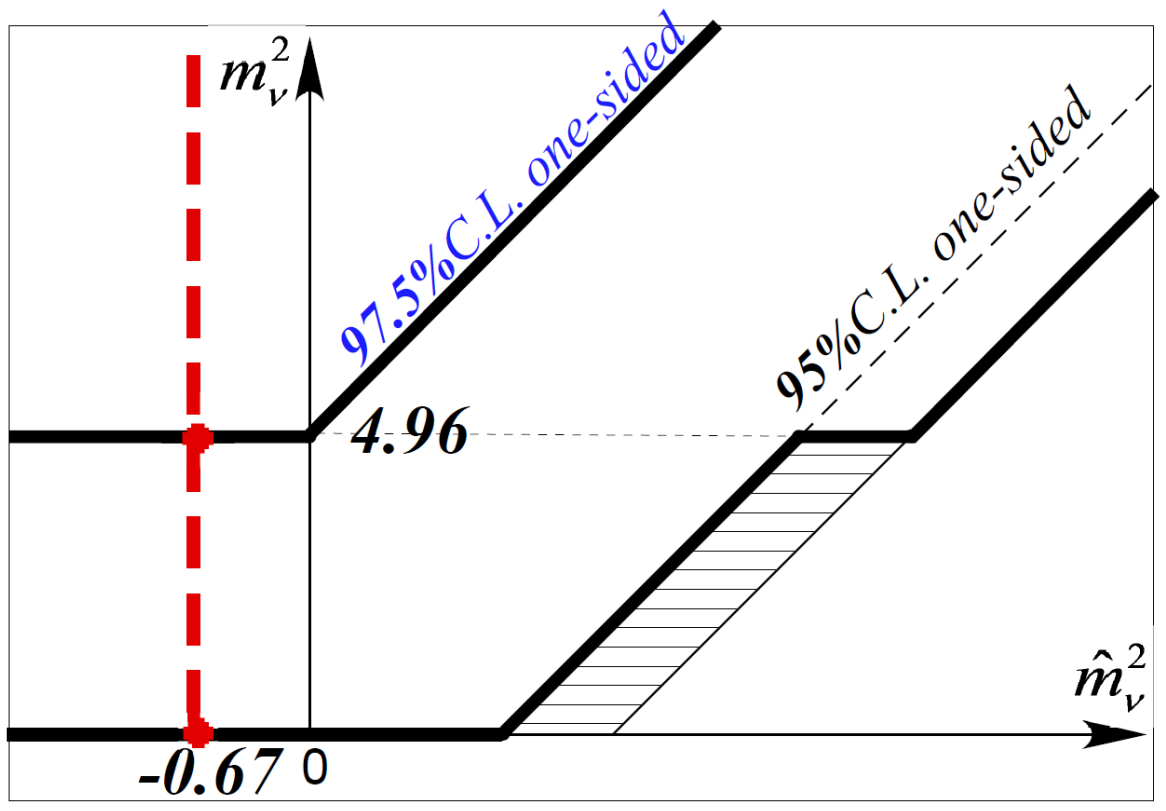

**Fig. 17.** Application of the confidence belt constructed via the method of sensitivity limit. The fat black lines correspond to the 95% confidence belt (see also Fig. 6). The confidence interval for the experimental estimate $\tilde{m}_\nu^2 = -0.67\ eV^2$ is determined by the ordinates of the intersections of the confidence belt and the dashed line $\tilde{m}_\nu^2 = -0.67\ eV^2$.

## 10. Conclusions

We have demonstrated that the problem of inclusion of a priori information about the estimated parameters into the construction of confidence intervals admits of a fully correct solution. We call the solution the method of sensitivity limit, since the solution partially resembles the well-known recipe known as sensitivity limit.

The solution has a transparent and therefore convincing basis, and is convenient for practical use (see the analytical descriptions of the confidence belts in Sections 6,7 and 8, and also Fig. 6, 8, 14 and 15).

Let us list the main features of the considered construction.

*Continuous distributions*

The confidence intervals constructed on base of the estimator that takes into account the a priori information about the bounded parameter of a continuous distribution possess the following properties (for the symmetric interval as well as for the best upper limit):

— The estimation is robust: it provides correct intervals for the non-physical values of the estimator, i.e. for $\hat{\theta} < 0$.

— The upper boundary of the interval remains the same for the physically allowed values of $\hat{\theta}$ both for the modified and non-modified estimators. The upper boundary is the lowest possible for the given confidence level $\beta$ in the case of the best upper limit construction. In the case of a symmetric interval for the physical values of $\hat{\theta}$ the upper boundary of the interval falls below the upper boundary of the construction [6].

— The lower boundary of the interval move away from the axis at the earliest possible (for the given confidence level) point $\theta_C$ (Fig. 6, 14). Besides the lower boundary appears to be maximal for the given confidence level within the segments $\hat{\theta}_C \le \hat{\theta} \le \hat{\theta}_E$ (Fig. 6), $\hat{\theta}_C \le \hat{\theta} \le \hat{\theta}_D$ (Fig. 14).

— The calculation of the confidence intervals does not require any sophisticated algorithms or tables apart from the standard procedures of constructing the confidence belts for a given confidence level.

*Discrete distributions*

Using the information about the background while choosing the estimator for a parameter of a discrete distribution one obtains the confidence intervals with the following properties:

— The described construction of the confidence intervals solves the problem of less than the background number of events, pointed out in [6]. The resulting system of confidence intervals immediately yields the upper limit for the estimated parameter in the region where the number of events is less than the mean of the background. The upper limit does not depend on the particular value of $n$ (as in the area $n \le 3$ in Fig. 8).

— The lower boundary contains a segment, analogous to the area CEF in Fig. 6. As a result of the use of the a priori information the lower boundary moves away from the axis $n$ at the earliest point possible.

— Due to the ambiguity in the definition of the confidence intervals for the discrete distributions (Eq. (10)) there are various modifications of the construction of confidence intervals (with fixed lower or upper boundary, the Sterne, Crow and Gardner construction and their combinations). The specific choice depends on the particular situation, for instance, on the necessity to obtain more reliable upper or lower constraints on the parameters.

The best upper limit for the unknown signal $\mu$ with the Poisson background with the mean value $b$ constructed on base of the estimator (20) possesses the following features:

— Any measured number of events $n \le b$ yields the same confidence interval. It solves the problem of less than the background number of events.

— The amount of probability contained in the acceptance region is close to the required confidence level, the overcoverage is not considerable (though one can not avoid it entirely in the case of discrete distributions). The overcoverage (in terms of [6]) is an undesirable feature of the various other constructions [19], [7].

— The lower boundary of the confidence intervals move away from the axis *n* (becomes non-zero) at the lowest possible value of *n*. Thus, one can not only obtain the best (the most stringent) upper limit for the parameter but also has an opportunity to detect the signal (the non-zero lower boundary can be interpreted as a signature of the non-zero signal) for rather small measured number of events.

— Similarly to the continuous distributions, there is an area below the lower boundary that represents a pure gain from the use of the a priori information about the background.

It should be remembered that an essential goal of data processing is that the magnitude of the resulting quantities should represent the required information as directly as possible. Unlike the approaches of Feldman and Cousins and Cowen et al., the method of sensitivity limit achieves that goal: its confidence intervals obtained in different experiments can be compared directly.

**Acknowledgements.** The authors thank the members of the Department of Experimental Physics of INR RAS for discussions. Special thanks are due to A.I. Berlev for his invariable interest in this work. We also thank A.S. Barabash for critical remarks on the software.

This work was supported in part by Russian Foundation for Basic Research grant 14-02-31055 and Russian Federation grant NS-2835.2014.2.

## APPENDIX

## Table 1

90% C.L. confidence intervals with account of the a priori information about the parameter $\mu$ with the known mean value of the background $b$=0..10 and the measured number of events $n_0 = (0..20)$. (in analogy to Tables II-IX in [6])

| $n_0 \setminus b$ | 0.0 | 0.5 | 1.0 | 1.5 | 2.0 | 2.5 | 3.0 | 3.5 |
|---|---|---|---|---|---|---|---|---|
| 0 | 0.0, 3.0. | 0.0, 2.5. | 0.0, 3.75. | 0.0, 3.25. | 0.0, 4.3. | 0.0, 3.8. | 0.0, 4.76. | 0.0, 4.26. |
| 1 | 0.11, 4.75. | 0.0, 4.25. | 0.0, 3.75. | 0.0, 3.25. | 0.0, 4.3. | 0.0, 3.8. | 0.0, 4.76. | 0.0, 4.26. |
| 2 | 0.54, 6.3. | 0.04,5.8. | 0.0, 5.3. | 0.0, 4.8. | 0.0, 4.3. | 0.0, 3.8. | 0.0, 4.76. | 0.0, 4.26. |
| 3 | 1.11, 7.76. | 0.61,7.26. | 0.11,6.76. | 0.0, 6.26. | 0.0,5.76. | 0.0, 5.26. | 0.0, 4.76. | 0.0, 4.26. |
| 4 | 1.75, 9.16. | 1.25,8.66. | 0.75,8.16. | 0.25, 7.66. | 0.0,7.16. | 0.0, 6.66. | 0.0, 6.16. | 0.0, 5.66. |
| 5 | 2.44, 10.52. | 1.94,10.02. | 1.44,9.52. | 0.94, 9.02. | 0.44,8.52. | 0.0, 8.02. | 0.0, 7.52. | 0.0, 7.02. |
| 6 | 3.0, 11.85. | 2.5,11.35. | 2.16,10.85. | 1.66, 10.35. | 1.16,9.85. | 0.66,9.35. | 0.16, 8.85. | 0.0, 8.35. |
| 7 | 3.29, 13.15. | 2.79,12.65. | 2.9,12.15. | 2.4, 11.65. | 1.9,11.15. | 1.4,10.65. | 0.9, 10.15. | 0.4, 9.65. |
| 8 | 3.99, 14.44. | 3.49,13.94. | 3.66,13.44. | 3.16, 12.94. | 2.66,12.44. | 2.16,11.94. | 1.66, 11.44. | 1.16, 10.94. |
| 9 | 4.7, 15.71. | 4.2,15.21. | 3.75,14.71. | 3.25, 14.21. | 3.44,13.71. | 2.94,13.21. | 2.44, 12.71. | 1.94, 12.21. |
| 10 | 5.43, 16.97. | 4.93,16.47. | 4.43,15.97. | 3.93, 15.47. | 4.23,14.97. | 3.73,14.47. | 3.23, 13.97. | 2.73, 13.47. |
| 11 | 6.17, 18.21. | 5.67,17.71. | 5.17,17.21. | 4.67, 16.71. | 4.3,16.21. | 3.8,15.71. | 4.03, 15.21. | 3.53, 14.71. |
| 12 | 6.93, 19.45. | 6.43,18.95. | 5.93,18.45. | 5.43, 17.95. | 4.93,17.45. | 4.43,16.95. | 4.76, 16.45. | 4.26, 15.95. |
| 13 | 7.69, 20.67. | 7.19,20.17. | 6.69,19.67. | 6.19, 19.17. | 5.69,18.67. | 5.19,18.17. | 4.76, 17.67. | 4.26, 17.17. |
| 14 | 8.47, 21.89. | 7.97,21.39. | 7.47,20.89. | 6.97, 20.39. | 6.47,19.89. | 5.97,19.39. | 5.47, 18.89. | 4.97, 18.39. |
| 15 | 9.25, 23.1. | 8.75,22.6. | 8.25,22.1. | 7.75, 21.6. | 7.25,21.1. | 6.75,20.6. | 6.25, 20.1. | 5.75, 19.6. |
| 16 | 10.04,24.31. | 9.54,23.81. | 9.04,23.31. | 8.54, 22.81. | 8.04,22.31. | 7.54,21.81. | 7.04, 21.31. | 6.54, 20.81. |
| 17 | 10.84,25.5. | 10.34,25.0. | 9.84,24.5. | 9.34, 24.0. | 8.84,23.5. | 8.34,23.0. | 7.84, 22.5. | 7.34, 22.0. |
| 18 | 11.64,26.7. | 11.14,26.2. | 10.64,25.7. | 10.14,25.2. | 9.64,24.7. | 9.14,24.2. | 8.64, 23.7. | 8.14, 23.2. |
| 19 | 12.45,27.88. | 11.95,27.38. | 11.45,26.88. | 10.95,26.38. | 10.45,25.88. | 9.95,25.38. | 9.45, 24.88. | 8.95, 24.38. |
| 20 | 13.26,29.07. | 12.76,28.57. | 12.26,28.07. | 11.76,27.57. | 11.26,27.07. | 10.76,26.57. | 10.26,26.07. | 9.76, 25.57. |

| $n_0 \setminus b$ | 4.0 | 4.5 | 5.0 | 5.5 | 6.0 | 6.5 | 7.0 | 7.5 |
|---|---|---|---|---|---|---|---|---|
| 0 | 0.0, 5.16. | 0.0, 4.66. | 0.0, 5.52. | 0.0, 5.02. | 0.0,5.85. | 0.0, 5.35. | 0.0, 6.15. | 0.0, 5.65. |
| 1 | 0.0, 5.16. | 0.0, 4.66. | 0.0, 5.52. | 0.0, 5.02. | 0.0,5.85. | 0.0, 5.35. | 0.0, 6.15. | 0.0, 5.65. |
| 2 | 0.0, 5.16. | 0.0, 4.66. | 0.0, 5.52. | 0.0, 5.02. | 0.0,5.85. | 0.0, 5.35. | 0.0, 6.15. | 0.0, 5.65. |
| 3 | 0.0, 5.16. | 0.0, 4.66. | 0.0, 5.52. | 0.0, 5.02. | 0.0,5.85. | 0.0, 5.35. | 0.0, 6.15. | 0.0, 5.65. |
| 4 | 0.0, 5.16. | 0.0, 4.66. | 0.0, 5.52. | 0.0, 5.02. | 0.0,5.85. | 0.0, 5.35. | 0.0, 6.15. | 0.0, 5.65. |
| 5 | 0.0, 6.52. | 0.0, 6.02. | 0.0, 5.52. | 0.0, 5.02. | 0.0,5.85. | 0.0, 5.35. | 0.0, 6.15. | 0.0, 5.65. |
| 6 | 0.0, 7.85. | 0.0, 7.35. | 0.0, 6.85. | 0.0, 6.35. | 0.0,5.85. | 0.0, 5.35. | 0.0, 6.15. | 0.0, 5.65. |
| 7 | 0.0, 9.15. | 0.0, 8.65. | 0.0, 8.15. | 0.0, 7.65. | 0.0,7.15. | 0.0, 6.65. | 0.0, 6.15. | 0.0, 5.65. |
| 8 | 0.66, 10.44. | 0.16,9.94. | 0.0, 9.44. | 0.0, 8.94. | 0.0,8.44. | 0.0, 7.94. | 0.0, 7.44. | 0.0, 6.94. |
| 9 | 1.44, 11.71. | 0.94,11.21. | 0.44,10.71. | 0.0, 10.21. | 0.0,9.71. | 0.0, 9.21. | 0.0, 8.71. | 0.0, 8.21. |
| 10 | 2.23, 12.97. | 1.73,12.47. | 1.23,11.97. | 0.73, 11.47. | 0.23,10.97. | 0.0,10.47. | 0.0, 9.97. | 0.0, 9.47. |
| 11 | 3.03, 14.21. | 2.53,13.71. | 2.03,13.21. | 1.53, 12.71. | 1.03,12.21. | 0.53,11.71. | 0.03, 11.21. | 0.0, 10.71. |
| 12 | 3.83, 15.45. | 3.33,14.95. | 2.83,14.45. | 2.33, 13.95. | 1.83,13.45. | 1.33,12.95. | 0.83, 12.45. | 0.33, 11.95. |
| 13 | 4.65, 16.67. | 4.15,16.17. | 3.65,15.67. | 3.15, 15.17. | 2.65,14.67. | 2.15,14.17. | 1.65, 13.67. | 1.15, 13.17. |
| 14 | 5.16, 17.89. | 4.66,17.39. | 4.47,16.89. | 3.97, 16.39. | 3.47,15.89. | 2.97,15.39. | 2.47, 14.89. | 1.97, 14.39. |
| 15 | 5.25, 19.1. | 4.75,18.6. | 5.3, 18.1. | 4.8, 17.6. | 4.3,17.1. | 3.8, 16.6. | 3.3, 16.1. | 2.8, 15.6. |
| 16 | 6.04, 20.31. | 5.54,19.81. | 5.52,19.31. | 5.02, 18.81. | 5.14,18.31. | 4.64,17.81. | 4.14, 17.31. | 3.64, 16.81. |
| 17 | 6.84, 21.5. | 6.34,21.0. | 5.84,20.5. | 5.34, 20.0. | 5.85,19.5. | 5.35,19.0. | 4.98, 18.5. | 4.48, 18.0. |
| 18 | 7.64, 22.7. | 7.14,22.2. | 6.64,21.7. | 6.14, 21.2. | 5.85,20.7. | 5.35,20.2. | 5.83, 19.7. | 5.33, 19.2. |
| 19 | 8.45, 23.88. | 7.95,23.38. | 7.45,22.88. | 6.95, 22.38. | 6.45,21.88. | 5.95,21.38. | 6.15, 20.88. | 5.65, 20.38. |
| 20 | 9.26, 25.07. | 8.76,24.57. | 8.26,24.07. | 7.76, 23.57. | 7.26,23.07. | 6.76,22.57. | 6.26, 22.07. | 5.76, 21.57. |

| $n_0 \backslash b$ | 8.0 | 8.5 | 9.0 | 9.5 | 10.0 |
|---|---|---|---|---|---|
| 0 | 0.0, 6.44. | 0.0, 5.94. | 0.0, 6.71. | 0.0, 6.21. | 0.0,6.97. |
| 1 | 0.0, 6.44. | 0.0, 5.94. | 0.0, 6.71. | 0.0, 6.21. | 0.0,6.97. |
| 2 | 0.0, 6.44. | 0.0, 5.94. | 0.0, 6.71. | 0.0, 6.21. | 0.0,6.97. |
| 3 | 0.0, 6.44. | 0.0, 5.94. | 0.0, 6.71. | 0.0, 6.21. | 0.0,6.97. |
| 4 | 0.0, 6.44. | 0.0, 5.94. | 0.0, 6.71. | 0.0, 6.21. | 0.0,6.97. |
| 5 | 0.0, 6.44. | 0.0, 5.94. | 0.0, 6.71. | 0.0, 6.21. | 0.0,6.97. |
| 6 | 0.0, 6.44. | 0.0, 5.94. | 0.0, 6.71. | 0.0, 6.21. | 0.0,6.97. |
| 7 | 0.0, 6.44. | 0.0, 5.94. | 0.0, 6.71. | 0.0, 6.21. | 0.0,6.97. |
| 8 | 0.0, 6.44. | 0.0, 5.94. | 0.0, 6.71. | 0.0, 6.21. | 0.0,6.97. |
| 9 | 0.0, 7.71. | 0.0, 7.21. | 0.0, 6.71. | 0.0, 6.21. | 0.0,6.97. |
| 10 | 0.0, 8.97. | 0.0, 8.47. | 0.0, 7.97. | 0.0, 7.47. | 0.0,6.97. |
| 11 | 0.0, 10.21. | 0.0, 9.71. | 0.0, 9.21. | 0.0, 8.71. | 0.0,8.21. |
| 12 | 0.0, 11.45. | 0.0,10.95. | 0.0,10.45. | 0.0, 9.95. | 0.0,9.45. |
| 13 | 0.65, 12.67. | 0.15,12.17. | 0.0,11.67. | 0.0, 11.17. | 0.0,10.67. |
| 14 | 1.47, 13.89. | 0.97,13.39. | 0.47,12.89. | 0.0, 12.39. | 0.0,11.89. |
| 15 | 2.3, 15.1. | 1.8, 14.6. | 1.3, 14.1. | 0.8, 13.6. | 0.3,13.1. |
| 16 | 3.14, 16.31. | 2.64,15.81. | 2.14,15.31. | 1.64, 14.81. | 1.14,14.31. |
| 17 | 3.98, 17.5. | 3.48,17.0. | 2.98,16.5. | 2.48, 16.0. | 1.98,15.5. |
| 18 | 4.83, 18.7. | 4.33,18.2. | 3.83,17.7. | 3.33, 17.2. | 2.83,16.7. |
| 19 | 5.68, 19.88. | 5.18,19.38. | 4.68,18.88. | 4.18, 18.38. | 3.68,17.88. |
| 20 | 6.44, 21.07. | 5.94,20.57. | 5.53,20.07. | 5.03, 19.57. | 4.53,19.07. |